\documentclass[a4paper,11pt]{article}

\usepackage{textcomp}
\usepackage{natbib}
\usepackage{authblk}
\usepackage{eqnarray,amsmath}
\usepackage{amssymb}
\usepackage{mathtools}
\usepackage{graphicx}
\usepackage{bbm}
\usepackage{algorithm}
\usepackage{algorithmic}
\usepackage{subfigure}
\usepackage{bm}
\usepackage{url}
\usepackage{lineno}
\usepackage[margin=3cm]{geometry}
\usepackage{epstopdf}
\usepackage{setspace}

\DeclarePairedDelimiterX{\infdivx}[2]{(}{)}{%
	#1\;\delimsize\|\;#2%
}

\title{\bf Sequential Experimental Design for Predator-Prey Functional Response Experiments}

\date{}

\author[1,2]{Hayden~Moffat \footnote{e-mail: hayden.moffat@hdr.qut.edu.au}}
\author[1,3]{Markus~Hainy}
\author[4,5,6]{Nikos~E.~Papanikolaou}
\author[1,2]{Christopher~Drovandi}

\affil[1]{School of Mathematical Sciences, Queensland University of Technology, Brisbane, Australia}
\affil[2]{ARC Centre of Excellence for Mathematical and Statistical Frontiers (ACEMS), Queensland University of Technology, Brisbane, Australia}
\affil[3]{Institute of Applied Statistics, Johannes Kepler University, Linz, Austria}
\affil[4]{Directorate of Plant Protection, Greek Ministry of Rural Development and Food, Athens, Greece}
\affil[5]{Laboratory of Agricultural Zoology and Entomology, Agricultural University of Athens, Greece}
\affil[6]{Benaki Phytopathological Institute, Athens, Greece}

\begin{document}
	
	\newcommand{\vect}[1]{\boldsymbol{#1}}

	\setlength{\parindent}{0pc}
	\setlength{\parskip}{1ex}
	
	\maketitle
	\begin{abstract}
		Understanding functional response within a predator-prey dynamic is a cornerstone for many quantitative ecological studies.  Over the past 60 years, the methodology for modelling functional response has gradually transitioned from the classic mechanistic models to more statistically oriented models. To obtain inferences on these statistical models, a substantial number of experiments need to be conducted. The obvious disadvantages of collecting this volume of data include cost, time and the sacrificing of animals.  Therefore, optimally designed experiments are useful as they may reduce the total number of experimental runs required to attain the same statistical results. In this paper, we develop the first sequential experimental design method for predator-prey functional response experiments.  To make inferences on the parameters in each of the statistical models we consider, we use sequential Monte Carlo, which is computationally efficient and facilitates convenient estimation of important utility functions. It provides coverage of experimental goals including parameter estimation, model discrimination as well as a combination of these. The results of our simulation study illustrate that for predator-prey functional response experiments sequential design outperforms static design for our experimental goals. R code for implementing the methodology is available via \url{https://github.com/haydenmoffat/sequential_design_for_predator_prey_experiments}.

	\end{abstract}
	\noindent
	{\it Keywords:} Optimal experimental design; Mutual information; Sequential Monte Carlo; Model discrimination; Total entropy
	
	\newpage
	
	\section{Introduction}
	\label{sec:intro}
	
	The term functional response refers to the number of prey consumed per predator as a function of prey density \citep{solomon1949natural}. Predators\textquotesingle~ feeding behaviour can be classified according to the type of the functional response. In this task, when the consumption rate increases linearly with prey density up to a threshold level at which it remains constant, we speak of type I functional response, which is exclusive to filter feeders \citep{jeschke2004consumer}. In type II functional response, the consumption rate continuously increases at a decelerating rate, whereas in type III it follows a sigmoid curve \citep{holling1959b}. Both type II and III functional responses reach a plateau at high prey densities and seem to prevail in nature \citep{jeschke2004consumer, sarnelle2008type}.
	
	Mathematical modelling and statistical analysis of functional response plays a crucial role in ecology as it enables us to gain a better understanding of the predator-prey interactions. Biological invasions, species extinction, biological control practices, as well as the management of ecosystems are strongly related to predators\textquotesingle~ functional response \citep{smith1973stability, papanikolaou2011functional, papanikolaou2014digestion, dick2014advancing}. It has been shown that type II functional response destabilise predator-prey dynamics, whereas at low prey densities type III functional response acts as a stabilising factor \citep{oaten1975functional}. Thus, describing a predator-prey system in such a quantitative manner allows for more accurate prediction and simulations. 
	
	Predator-prey functional response experiments are set up so that a single predator (or multiple predators) has access to fixed numbers of prey for a given period of time. The number of prey that are attacked out of the total that the predator has access to in that given time period is recorded. In order to gain inferential information from predator-prey functional response experiments, several trials need to be conducted. The obvious disadvantages of collecting this inordinate volume of data and conducting these experiments include cost, time and the sacrificing of animals. Consequently, optimal experimental design has become beneficial to behavioural ecologists to reduce the number of experimental trials that need to be run. The experimental design involves optimising a particular measure for an experimental purpose or goal. 	
	
	The current literature for optimal experimental design for functional response models is scarce with one paper by \citet{zhang2018optimal}. The approach of \citet{zhang2018optimal} only considers static designs, which requires selecting the number of prey available to the predator(s) for each experiment in the study prior to any experimentation. If there is little prior information on functional response model parameters, then optimal designs may be inefficient. Furthermore, \citet{zhang2018optimal} consider optimal designs for the purpose of precise frequentist parameter estimation for a single, assumed true, model. However, \citet{Pap_etal16} demonstrate that there can be significant uncertainty in which functional response model might be responsible for data generation. Therefore, the ability to acknowledge model structure uncertainty and the ability to use optimal design to help discriminate between the models is highly desirable. 
	
	In this paper we develop the first sequential experimental design approach for predator-prey functional response experiments. Unlike the static design framework used by \citet{zhang2018optimal}, the sequential design set-up allows practitioners to update their information about model structure and parameter values as observations are collected sequentially. In the optimal design context, this is important as this additional information can lead to more efficient design choices for future observations. Moreover, in contrast to \citet{zhang2018optimal}, our approach explicitly accommodates uncertainty in the model structure. We consider optimal experimental design utility functions for the purpose of parameter estimation and/or model discrimination. Our sequential design methodology uses the sequential Monte Carlo (SMC) approach of \citet{drovandi2013sequential} and \citet{DrovandiEtAlDesign2012}, which is computationally efficient and permits convenient estimation of utility functions for parameter estimation or model discrimination. We demonstrate how the total entropy criterion of \citet{borth1975total}, a dual-purpose utility function for the goals of parameter estimation and model discrimination, is easily computed with our approach. 
	
	The rest of the paper is outlined as follows. Section \ref{sec:background} provides background information regarding the fundamentals of modelling functional response. A simple illustration of the sequential design process is included in Section \ref{sec:ill}. Section \ref{sec:SED} describes our sequential experimental design approach in more detail. Section \ref{sec:simstudy} outlines a simulation study that was conducted to demonstrate and quantify the benefits of using the methodology proposed in the paper. This simulation study enables us to gain insight into examples of a predator-prey interaction while also evaluating the performance of the algorithm and of resulting designs. The paper is then concluded in Section \ref{sec:diss} with a discussion of the simulation study results, the limitations of our approach and possible future work.

	\section{Background on Functional Response Models and Predator-Prey Functional Response Experiments}
	\label{sec:background}
	
	Among current behavioural ecologists, the mechanistic equations developed by \citet{holling1959b} are favoured when modelling functional response for predator-prey interactions (see, for example, \citealp{beddington1975mutual, okuyama2012a}). The preference for these models stems from their simple structure, where parameters can be easily translated to physical phenomena such as consumption rate and handling time. The simplest of Holling\textquotesingle s equations is often referred to as the disc equation or Holling\textquotesingle s type II functional response model. Holling\textquotesingle s type II model is given by the ordinary differential equation,
	
	\begin{align}
	\frac{dN}{d\tau} = -~\frac{aN}{1+a T_h N}. 
	\label{eq:holling_discT2}
	\end{align}

	The parameters $a$ and $T_{h}$ represent the attack rate, i.e. the per capita prey consumption at low prey densities, and the handling time, i.e. the time a predator spends subduing, pursuing and eating a prey item, respectively. $N$ denotes the prey density in a given area, and is a function of $\tau$ (time). Holling\textquotesingle s modelling approach for type II illustrates a functional response curve where the consumption rate increases with prey density at a decelerating rate, until it reaches a plateau/constant consumption rate. An extension of the disc equation is Holling\textquotesingle s type III model \citep{holling1959b}. The type III functional response model describes situations in which the functional response curve forms a characteristic “S” shape. That is, the consumption rate accelerates at low prey densities, decelerates at high prey densities and then reaches a plateau/constant consumption rate. Holling\textquotesingle s type III model is given by 
	
	\begin{align}
	\frac{dN}{d\tau} = - ~ \frac{a N^{2}}{1+a T_h N^{2}}. 
	\label{eq:holling_discT3}
	\end{align}
	
	Although there are more complex prey-dependent functional response models in the current literature (see, for example, \citealp{Jeschke2002}), this paper will solely focus on the implementation of Holling\textquotesingle s type II and type III models due to their popularity and simplicity. However, our method can easily accommodate any functional response model.
	
\begin{figure}[H]
		\centering
		\includegraphics[height=0.4\textheight,keepaspectratio]{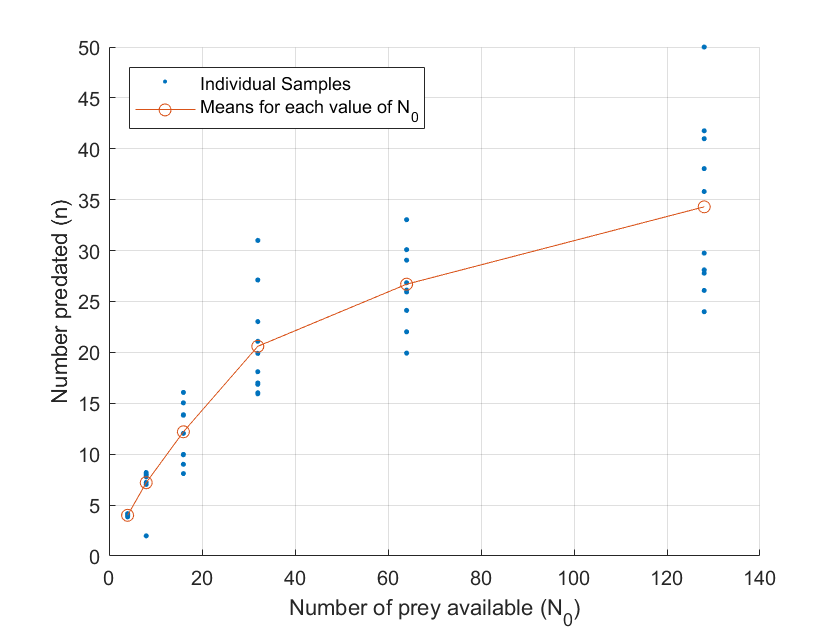}
		\caption{An example predator-prey functional response dataset from  \citet{Pap_etal16}.}
		\label{fig:example}
	\end{figure}

	The primary interest in \eqref{eq:holling_discT2} and \eqref{eq:holling_discT3} is on the parameters $T_{h}$ and $a$. To obtain more information on a predator-prey interaction, particularly inferences on the related parameters, experimental data with varying initial prey densities are collected. Figure \ref{fig:example} shows an example of predator-prey functional response data collected from an experiment conducted by \citet{Pap_etal16}. 
	
	For the study, $I$ independent runs of the predator-prey system are conducted. The initial prey density at run $i$ is denoted by $N_{0,i}$ for $i = 1,\ldots,I$. This variable is controlled by the experimenter. At each run, the number of prey consumed in a fixed time period (in hours), $\tau$, denoted by $n_{i}(\tau)$, is observed and used as the response variable. In this paper, we link Holling\textquotesingle s type II and type III models to probabilistic models to help account for uncertainty in the observational data. We link them in such a way so that solutions to mechanistic models are used to determine the expected proportions of prey eaten in the probabilistic models we consider.
	
	We define $n(\tau)$ to be the number of prey consumed/eaten for a single experiment that has a fixed time period of $\tau$. 	Given that $\tau$ is usually fixed across experiments, we write $n = n(\tau)$ for notational simplicity. We consider two possible distributions for $n$. Since for any fixed time period $\tau$, each of the $N_{0}$ prey is either dead or alive, a binomial distribution might be a reasonable assumption. Alternatively, in the case where the data seems to indicate overdispersion, which often arises in predator-prey functional response data \citep{trexler1988can}, the beta-binomial distribution may be more appropriate to describe the distribution of $n$. \citet{fenlon2006modelling} and \citet{zhang2018optimal} use the beta-binomial distribution to capture the variability of the data in a similar context.
	
	For the case where the number of prey consumed is modelled by the binomial distribution, we have a set-up that is similar to that of \citet{Pap_etal16}:
	
	\begin{align*}
	n \sim \mbox{Binom}(N_{0}, p_{\tau}),\\
	p_{\tau} = \frac{N_{0} - N_{\tau}}{N_{0}},
	\end{align*}
	
	where $p_{\tau}$ is the probability that a single prey has been consumed by time $\tau$ and $N_{\tau}$ is the solution of the differential equation of the type II or type III model. $p_{\tau}$ and $N_{\tau}$ both implicitly depend on the model parameters $a$ and $T_h$, but we do not explicitly write them here for notational convenience.
	
	For the beta-binomially distributed case, we have a set-up that is similar to \citet{zhang2018optimal}. The probability mass function for a single observation and the expected value are given by

	\begin{align}
	p(n;N_{0},\alpha,\beta) = \binom{N_{0}}{n} \frac{B(n + \alpha, N_{0}-n+\beta)}{B(\alpha, \beta)}, 
	\label{eq:betabinomial}
	\end{align}

	\begin{align}
	E[n] = \frac{N_{0} \alpha}{\alpha+ \beta},
	\label{eq:meanbetabin}
	\end{align}
	
	respectively. In \eqref{eq:betabinomial} and \eqref{eq:meanbetabin}, $\alpha$ and $\beta$ represent the two parameters of the beta-binomial distribution and $B(\cdot,\cdot)$ is the beta function. 
	
	To link the solutions of the mechanistic equations to the beta-binomial distribution, we re-parameterise the beta-binomial distribution in terms of the expected proportion, $p_{\tau}$, and over-dispersion parameter, $\lambda$, such that

	\begin{align*}
	p_{\tau} = \frac{\alpha}{\alpha+ \beta} &= \frac{N_{0} - N_{\tau} }{N_{0}} \mbox{ and} \\
	\lambda = \frac{1}{\alpha+ \beta}.
	\end{align*}

	Therefore, we have that
	
	\begin{align*}
	n \sim \mbox{BetaBinom}(N_{0}, p_{\tau}, \lambda).\\
	\end{align*}
	
	\newpage
	
		\section{Illustration of the Sequential Design Process}
		\label{sec:ill}
	
	To assist decision makers who need to optimise their predator-prey functional response experiments, we provide a simple illustration of the sequential design process. This section is less technical than other sections, with the purpose of making the methodology described in the paper more accessible to practitioners. Section \ref{sec:SED} provides a more technical overview of the methodology.	
	
	Consider a scenario where the aim of our predator-prey functional response experiments is efficient parameter estimation and model discrimination.  Our myopic sequential design approach involves running experiments one-at-a-time and using the results from the previous experiments to make more informed decisions about future experimentation. The set-up for our approach is simple and only requires the user to define prior probabilities for each candidate model and prior distributions for the parameters of each model. For this illustrative example, we design and conduct $4$ experiments. For illustrative purposes, we assume that the true model (model responsible for data generation) is a Holling\textquotesingle s type II beta-binomial model and its parameters are $a = 0.5$, $T_{h} = 0.7$ and $\lambda = 0.5$.

		\begin{figure}[H]
		\centering
		\subfigure[Iteration 1]{\includegraphics[height=0.22\textheight,keepaspectratio]{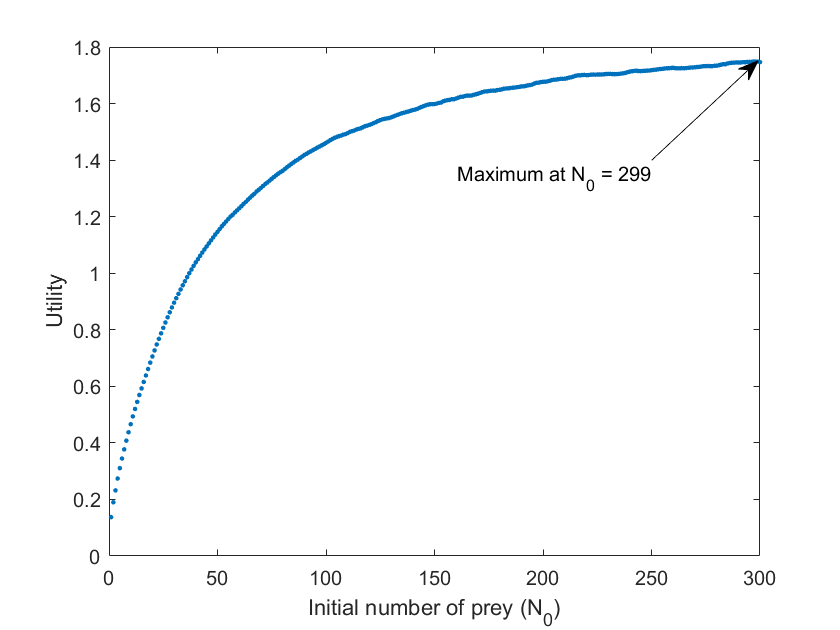}\label{figsub:u1}}
		\subfigure[Iteration 2]{\includegraphics[height=0.22\textheight,keepaspectratio]{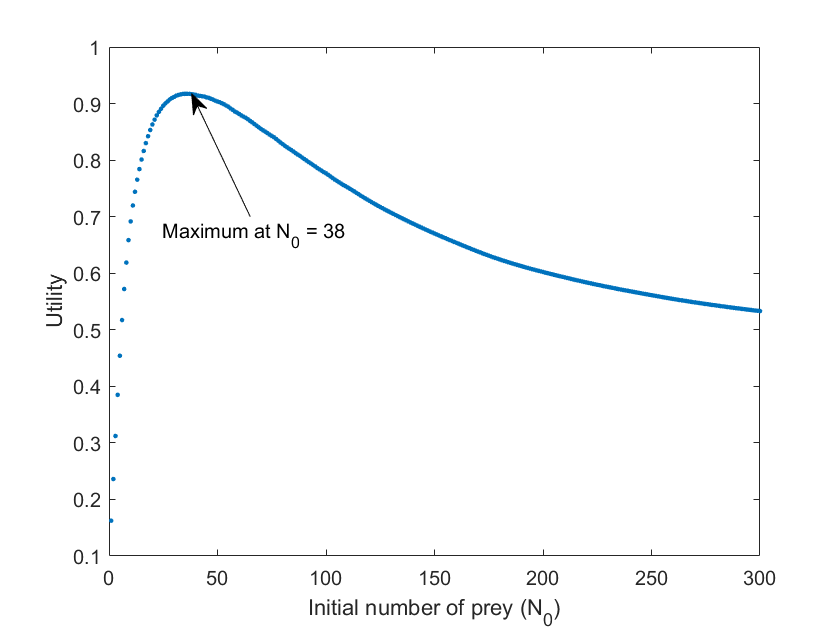}\label{figsub:u2}}
		\subfigure[Iteration 3]{\includegraphics[height=0.22\textheight,keepaspectratio]{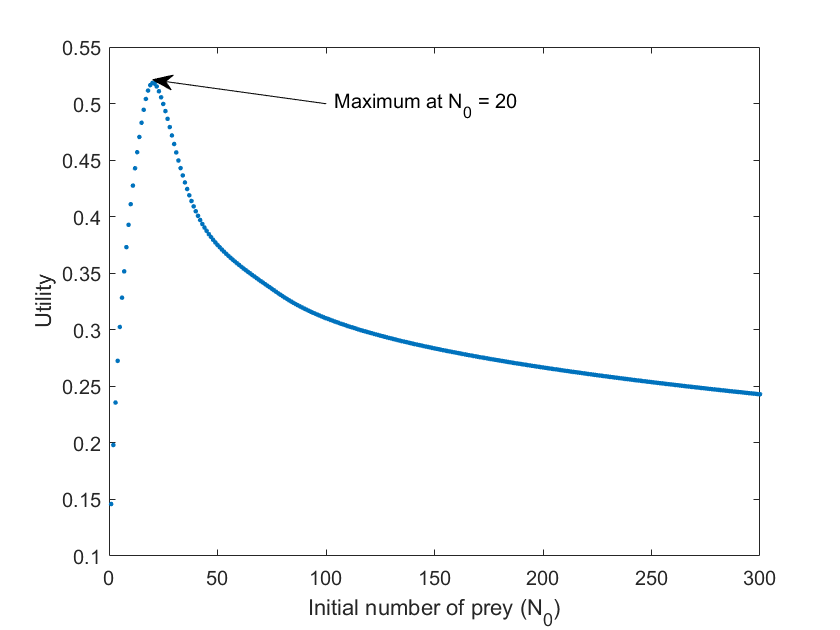}\label{figsub:u3}}
		\subfigure[Iteration 4]{\includegraphics[height=0.22\textheight,keepaspectratio]{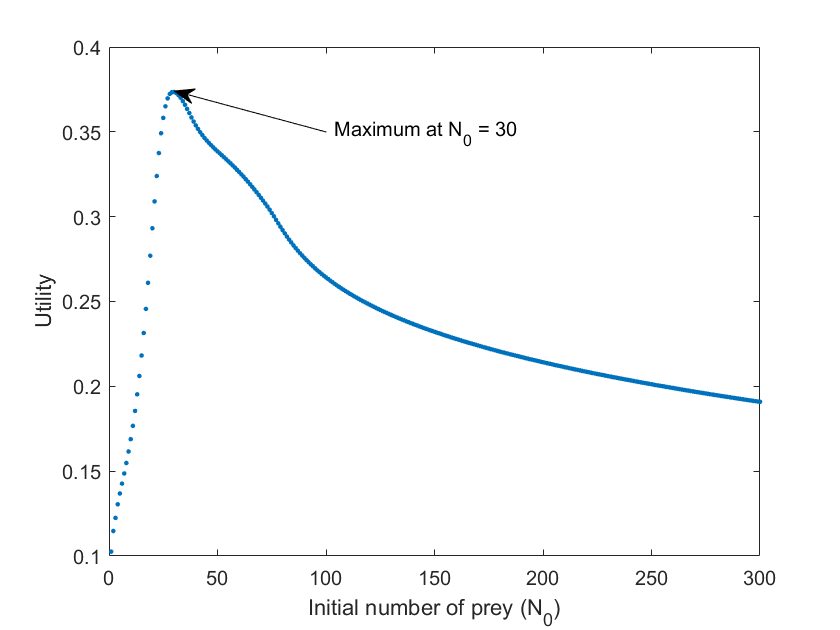}\label{figsub:u4}}
		\caption{The utility curve for each iteration of the sequential design process. The x-axis indicates the experimental design and the y-axis represents its corresponding utility value. The optimal design on each curve has been identified with an arrow.}
		\label{fig:u}
	\end{figure}
	
	\newpage
	
	Before we run an experiment, we need to make a decision regarding the  number of prey we initially make available to the predator.  The optimal initial prey density for the next experiment is determined by maximising a function that is related to our experimental objective. This function depends on our current knowledge of the parameters and model probabilities and is commonly referred to as a utility function. For more information on utility functions and how they capture the goals of an experiment see Section \ref{sec:utilities}. As mentioned earlier, the utility function for this illustration is designed to (on average) detect the preferred model and precisely estimate the corresponding model parameters as quickly as possible. Figure \ref{fig:u} shows the utility curve for each of the 4 experiments. The optimal initial prey density for the next experiment is identified on each curve. For this illustrative example, the initial prey densities for our 4 experiments are 299, 38, 20 and 30. The search space for the optimisation is simply all the feasible initial prey densities. We include all whole numbers from $1$ to $300$ in our search space for this example. 

\begin{figure}[H]
		\centering
		\includegraphics[height=0.3\textheight,keepaspectratio]{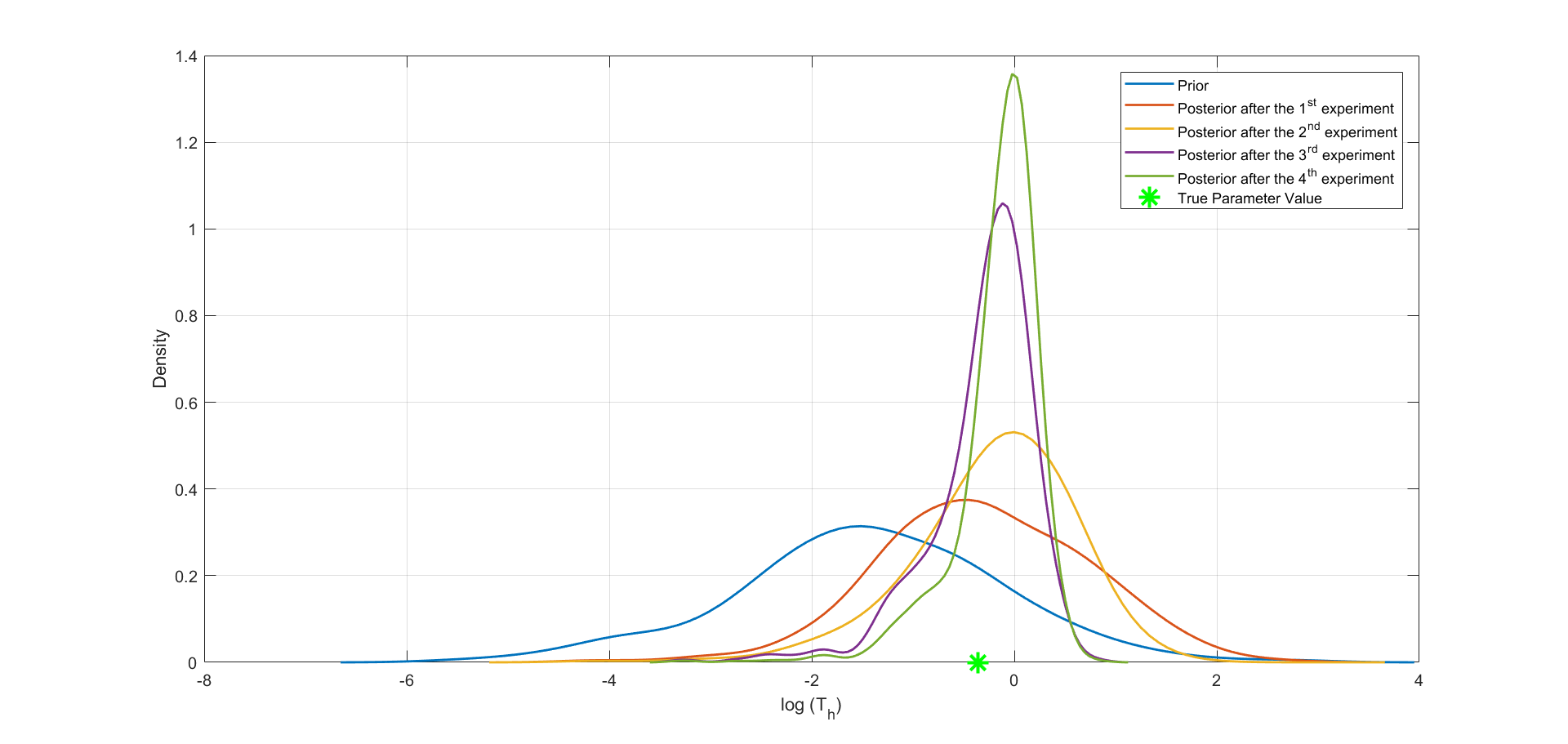}
		\caption{Marginal posterior distribution of $\log (T_{h})$ after $1$, $2$, $3$ and $4$ iterations of the sequential design process. The parameter $T_{h}$ is from the Holling\textquotesingle s type II beta-binomial functional response model. The marginal posteriors are compared to the prior and the true value of the parameter is displayed as a star.}
		\label{fig:posteriors}
	\end{figure}

	After the optimal initial prey density for the next experiment is identified, we conduct an experiment using that initial number of prey.
	We record the corresponding observation, which is the number of prey consumed, and update the posterior distribution. Here we update the posterior distribution using sequential Monte Carlo (see Section \ref{sec:SMC}) given its computational efficiency for updating posterior approximations.  However, other posterior sampling/approximation methods could be adopted here.  The posterior then becomes the prior for the next experiment. This process is repeated until a specified number of experiments have been run or a certain level of precision has been reached. Figure \ref{fig:posteriors} illustrates the evolution of the marginal posterior distribution of $\log (T_{h})$ over the 4 experiments. A continual increase in the precision of the parameter is clearly visible. The prior model probability of the true model is $0.25$. The posterior model probabilities of the true model after each of the 4 experiments are $0.45, 0.62, 0.52,$ and $0.47$ (ordered from the $1^{st}$ experiment to the $4^{th}$ experiment). Although the true model probability is fluctuating, the probability will tend to 1 with the collection of more data.

		\section{Sequential Experimental Design}
	\label{sec:SED}
	\subsection{General Notation}
	\label{sec:GN}
	
	The following section outlines the general notation that is used throughout the paper. Define $K$ to be the total number of candidate models. Define $M$ to be a random variable that indicates which model is responsible for data generation. $M$ can take the values $\left\{1,\ldots,K\right\}$. Let $\vect{y}_{1:i}$ denote a vector of all the observations up to experiment number $i$ and $\vect{d}_{1:i}$ represent a vector of all the selected design points up to experiment number $i$. The likelihood of observing $\vect{y}_{1:i}$ for model $m$ with a set of parameters $\vect{\theta}_{m}$ is denoted by $f(\vect{y}_{1:i}|M=m, \vect{\theta}_{m}, \vect{d}_{1:i})$. Denote $\pi_{0}(\vect{\theta}_{m}| M=m)$ to be our prior distribution, that is, our knowledge of the parameter $\vect{\theta}_{m}$ for model $m$ prior to the experiment. The posterior distribution of $\vect{\theta}_{m}$ for model $m$ after $i$ experiments is given by

	\begin{align*}
	\pi_{i}(\vect{\theta}_{m}|M=m, \vect{y}_{1:i}, \vect{d}_{1:i}) = \frac{f(\vect{y}_{1:i}|M=m, \vect{\theta}_{m}, \vect{d}_{1:i})~\pi_{0}(\vect{\theta}_{m}|M=m)}{Z_{m,i}},
	\end{align*}
	
	where $Z_{m,i}$ is the evidence for model $m$ and is given by the prior predictive probability of the observed data:

	\begin{align*}
	Z_{m,i} = f(\vect{y}_{1:i}|M=m, \vect{d}_{1:i}) = \int_{\vect{\theta}_{m}}f(\vect{y}_{1:i}|M=m, \vect{\theta}_{m}, \vect{d}_{1:i})~\pi_{0}(\vect{\theta}_{m}|M=m)~ \mathrm{d}\vect{\theta}_{m}.
	\end{align*}
	
	For the remainder of this article, $M=m$ will be referred to as just $m$ for simplicity. In the context of functional response models, $y_i = n_i$, $d_i = N_{0,i}$, $m$ refers to a particular functional response model and $\vect{\theta}_m$ its corresponding parameter; for example, $\vect{\theta}_m = (a,T_h,\lambda)$ for a beta-binomial type II or III functional response model.
	
	We now define the notation relevant to the sequential experimental design aspect of the paper. We denote the proposed design point as $d$ and a possible value of the response after we have observed $\vect{y}_{1:i}$ as $z$. Define $D$ to be a set of all the possible design points for a single observation and $S$ be a set of all the possible responses. The utility for the design point $d$ at observation $z$ and for model $m$ based on the current data is denoted $U(d,z, m| \vect{y}_{1:i}, \vect{d}_{1:i})$. The utility for the proposed design point, $U(d|\vect{y}_{1:i}, \vect{d}_{1:i})$, can be obtained by taking the expectation over the model and observation space. Section \ref{sec:utilities} outlines the specific utility functions used in this paper.
	
	When we collect a new observation, we can easily update the posterior, assuming independence among observations, by multiplying the current posterior by the likelihood of the next observation $(d,z)$:

	\begin{align}
	\pi_{i+1}(\vect{\theta}_{m}|m, \vect{y}_{1:i}, z, \vect{d}_{1:i},d) \propto f(z|m, \vect{\theta}_{m}, d) \pi_{i}(\vect{\theta}_{m}|m,\vect{y}_{1:i}, \vect{d}_{1:i}) \mbox{  for } i=0,...,I-1.
	\label{eq:seq}
	\end{align}

	\subsection{Sequential Optimal Design}	
	\label{sec:SOD}
	In this section we discuss the relevant theory necessary to understand the proposed sequential optimal experimental design algorithm. The algorithm itself is presented in Section \ref{sec:estimation}. Sequential experimental design involves the utilisation of previously collected data in conjunction with a utility function to improve future data collection. We collect data points one-at-a-time and make an informed decision on the next design point. This myopic approach to experimental design has many advantages over static designs. Sequential experimental designs are generally more efficient in the presence of parameter and model uncertainty (see, for example, \citealp{dror2008sequential}) and involve lower-dimensional design optimisation problems at each iteration.  Another benefit is that the sequential nature of the experimental design is well suited to SMC. SMC has several benefits which will be discussed later in Section \ref{sec:SMC}.  
	
	Optimal experimental design involves selecting design points such that the experimental goals are achieved in the minimum possible number of experimental runs.  In this paper we consider the experimental goals of parameter estimation, model discrimination and a combination of these.  The experimental goals can be captured by different utility functions which depend on the currently collected data.  Define $d^*$ to be the optimal design point for the next observation. We obtain the optimal design point by maximising the utility over the design space $D$: 
	
	\begin{align*}
	d^* = \mbox{arg }  \underset{d \epsilon D}{\mbox{max }} U(d|\vect{y}_{1:i}, \vect{d}_{1:i}).
	\end{align*}
	
	The utility of design point $d$, $U(d|\vect{y}_{1:i}, \vect{d}_{1:i})$, is determined by taking the expectation of the user-specified utility function, $U(d,z,m|\vect{y}_{1:i}, \vect{d}_{1:i})$,  over the response and model space:

	\begin{align}
	U(d|\vect{y}_{1:i},\vect{d}_{1:i}) &=
	\sum_{m=1}^{K}
	\pi_{i}(m|\vect{y}_{1:i},\vect{d}_{1:i})
	\sum_{z\epsilon S} 
	f(z|m,\vect{y}_{1:i},\vect{d}_{1:i},d)~
	U(d,z,m| \vect{y}_{1:i}, \vect{d}_{1:i}).
	\label{eq:utility}
	\end{align}
	
	The quantity $f(z|m,\vect{y}_{1:i},\vect{d}_{1:i},d)$ is the posterior predictive probability of a future observation and is given by

	\begin{align*}
	f(z|m,\vect{y}_{1:i},\vect{d}_{1:i},d) = 
	\int_{\vect{\theta}_{m}} f(z|m,\vect{\theta}_{m},d)~\pi_{i}(\vect{\theta}_{m}|m,\vect{y}_{1:i}, \vect{d}_{1:i})~\mathrm{d}\vect{\theta}_{m}.
	\end{align*}

	\subsection{Utility Functions}
	\label{sec:utilities}
	Selecting a utility function that adequately captures the goals of an experiment is an integral part of optimal experimental design. Our aim is to select design points in order to increase our certainty around the ``true model'' upon observation of the experimental outcomes. In this section, we outline the three utility functions used in our SMC algorithm, all of which correspond to a specific experimental goal.
	
	An important component of all the utilities is the Kullback-Leibler divergence (KLD) \citep{kullback1951information}. The KLD is an information-based measure of disparity between two distributions. In our case, the KLD represents the information gain on the true data generating process.

	\subsubsection{Parameter Estimation Utility}
	\label{sec:PE}
	If the objective of our sequential experimental design is to maximise the precision of model parameter posterior distributions, the KLD between the current and updated posterior distributions is a highly useful utility. For model $m$, the KLD between the current posterior, $\pi_{i}(\vect{\theta}_{m}|m,\vect{y}_{1:i},\vect{d}_{1:i})$, and the posterior based on the observation $z$ and the proposed design point $d$, $\pi_{i+1}(\vect{\theta}_{m}|m,\vect{y}_{1:i},z,\vect{d}_{1:i},d)$, is given by

	\begin{align}
	U(d,z,m|\vect{y}_{1:i}, \vect{d}_{1:i})= \int_{\vect{\theta}_{m}} \pi_{i+1}(\vect{\theta}_{m}|\vect{y}_{1:i},z,\vect{d}_{1:i}, d) ~ \log{\left(\frac{\pi_{i+1}(\vect{\theta}_{m}|\vect{y}_{1:i},z,\vect{d}_{1:i}, d)}{\pi_{i}(\vect{\theta}_{m}|\vect{y}_{1:i}, \vect{d}_{1:i})}\right)} ~ \mathrm{d}\vect{\theta}_{m},
	\label{eq:util_pe}
	\end{align}
	
	where the dependency of the current and updated posterior on $m$ is omitted for brevity. Equation \eqref{eq:util_pe} is simplified to,

	\begin{align}
	U(d,z,m|\vect{y}_{1:i}, \vect{d}_{1:i})
	= \int_{\vect{\theta}_{m}} \pi_{i+1}(\vect{\theta}_{m}|\vect{y}_{1:i},z,\vect{d}_{1:i}, d) \log{f(z|\vect{\theta}_{m},d)}~ \mathrm{d}\vect{\theta}_{m} - \log{\left(\frac{Z_{m,i}(d, z)}{Z_{m,i}}\right)},
	\label{eq:util_pe2}
	\end{align}
	
	where again the dependency of the current and updated posterior as well as the likelihood of $z$ on $m$ is omitted. The value $Z_{m,i}(d, z)$ represents the evidence at experiment number $i+1$ for model $m$ if we observe the response $z$ at the next design point $d$. 
	
	The utility given in \eqref{eq:util_pe2} will allow us to optimally design an experiment for the goal of parameter estimation for all of the $K$ candidate models. The utility for design point $d$, $U(d|\vect{y}_{1:i},\vect{d}_{1:i})$,  is given by substituting \eqref{eq:util_pe2} into \eqref{eq:utility}.
	
	\subsubsection{Model Discrimination Utility}
	\label{sec:MD}
	Alternatively, a model discrimination utility may be of interest. In this case, we use a utility which is based upon the mutual information between the model indicator, $m$, and the predicted observation, $z$. The mutual information is mathematically equivalent to the KLD between the joint distribution of $m$ and $z$ and the product of the marginal distributions of $m$ and $z$. This utility was initially suggested by \citet{box1967discrimination} and has been recently implemented by  \citet{DrovandiEtAlDesign2012}. The utility for model discrimination is given by

	\begin{align}
	U(d,z,m|\vect{y}_{1:i}, \vect{d}_{1:i}) =  \log \pi(m|\vect{y}_{1:i},z,\vect{d}_{1:i},d).
	\label{eq:util_md}
	\end{align}
	
	The utility for design point $d$, $U(d|\vect{y}_{1:i},\vect{d}_{1:i})$, is given by substituting \eqref{eq:util_md} into \eqref{eq:utility}.

	\subsubsection{Dual-Purpose Utility}
	\label{sec:DP}
	Similar to the design problems discussed by \citet{dette2001robust}, \citet{zen2004criterion} and \citet{senarathne2019bayesian}, we consider a dual-purpose experimental goal which combines parameter estimation and model discrimination using the total entropy criterion \citep{borth1975total}. Denote $U_{\mathrm{PE}}(d,z,m| \vect{y}_{1:i}, \vect{d}_{1:i})$ to be the parameter estimation utility from \eqref{eq:util_pe2} and $U_{\mathrm{MD}}(d,z,m|\vect{y}_{1:i}, \vect{d}_{1:i})$ denote the model discrimination utility from \eqref{eq:util_md}. The dual-purpose utility for design point $d$ is given by 

	\begin{align}
	U(d,z,m|\vect{y}_{1:i}, \vect{d}_{1:i}) = U_{\mathrm{PE}}(d,z,m| \vect{y}_{1:i}, \vect{d}_{1:i}) + U_{\mathrm{MD}}(d,z,m| \vect{y}_{1:i}, \vect{d}_{1:i}),
	\label{eq:util_dp}
	\end{align}
	
	which is purely the sum of the parameter estimation and model discrimination utilities. 
	
	Through the process of simplifying and removing terms which do not depend on the design point $d$, we arrive at a dual-purpose utility: 
	
	\begin{align*}
	U(d|\vect{y}_{1:i},\vect{d}_{1:i}) &= \sum_{m=1}^{K} \pi_{i}(m|\vect{y}_{1:i},\vect{d}_{1:i})
	\sum_{z\epsilon S}f(z|m,\vect{y}_{1:i},\vect{d}_{1:i},d) 
	\\&\int_{\vect{\theta}_{m}} \pi_{i+1}(\vect{\theta}_{m}|m,\vect{y}_{1:i},z,\vect{d}_{1:i}, d) \log{f(z|m,\vect{\theta}_{m},d)}~ \mathrm{d}\vect{\theta}_{m} \\&
	- \sum_{z\epsilon S} f(z|\vect{y}_{1:i}, \vect{d}_{1:i},d) \log f(z|\vect{y}_{1:i}, \vect{d}_{1:i},d).
	\end{align*}
	
	The posterior predictive distribution $f(z|\vect{y}_{1:i}, \vect{d}_{1:i},d)$  is determined by averaging $f(z | m, \vect{y}_{1:i}, \vect{d}_{1:i}, d)$ over all the models.
	
	Given the form of these utility functions, most of these utilities are analytically intractable and therefore must be estimated. Unfortunately, estimating these quantities is not a straightforward process.  SMC enables us to form particle approximations to a number of intractable integrals contained within utility functions. In addition, the form of our utility functions is convenient for estimation through SMC. Sections \ref{sec:SMC} and \ref{sec:estimation} discuss these approximations in greater depth.

	\subsection{Sequential Monte Carlo}
	\label{sec:SMC}
	
	SMC samplers are a useful tool for assessing parameter and model uncertainty when conducting sequential experimental design. The main advantage of using an SMC framework for sequential design is that after collecting an observation, it enables us to obtain efficient approximations to the posterior and other quantities of interest. Consequently, this allows us to efficiently obtain approximations of utility functions (see Section \ref{sec:estimation}) as data is collected and thus allows us to easily explore the design space, $D$, for the next optimal design point.
	
	SMC involves traversing a set of $J$ weighted samples (particles) for each of our $K$ models through a sequence of slowly evolving target distributions by iteratively conducting re-weighting, resampling and move steps. We denote the set of particles representing the target for model $m$ at experiment number $i$ to be $\left\{\vect{\theta}_{m,i}^{j}\right\}^{J}_{j=1}$ with the corresponding weights $\left\{W_{m,i}^{j}\right\}^{J}_{j=1}$. We denote the unnormalised and normalised weights for the $j^{th}$ particle of model $m$ at experiment number $i$ as $w^j_{m,i}$ and $W^j_{m,i}$, respectively. In this implementation of SMC, the sequence of distributions is formed through the process of data annealing. This process involves setting up a sequence of distributions by introducing data one-at-a-time to arrive at the posterior.

	Given a particular model, $m$, the sequence of targets is given by	
	
	\begin{align*}
	\pi_{i}(\vect{\theta}_{m}|m, \vect{y}_{1:i}, \vect{d}_{1:i}) \propto f(\vect{y}_{1:i}|m, \vect{\theta}_{m}, \vect{d}_{1:i})~\pi_{0}(\vect{\theta}_{m}|m) \mbox{  for }  i=1,...,I.
	\end{align*}
	
	After an observation is collected, we initially re-weight the particles to reflect the new target distribution, which in our case is the updated posterior distribution. This is conducted through the process of importance sampling where the unnormalised weights for our new target are given by
	
	\begin{align*}
	w^j_{m,i+1} = W^j_{m,i} f(y_{i+1}|m, \vect{\theta}_{m, i}^{j}, d_{i+1}).
	\end{align*}
	
	After the re-weighting process is completed and the weights are normalised, the weights tend to become more skewed. This leads to the reduction in the effective sample size (ESS).  The ESS is a measure of the efficiency of a particle set and refers to the number of independent samples (of equal weight) from the target distribution that the weighted sample is worth.  The ESS for model $m$ at experiment number $i$ can be estimated by
	
	\begin{align}
	\mbox{ESS}_{m,i} = \frac{1}{\sum_{j=1}^{J}(W_{m,i}^{j})^{2}}.
	\label{eq:ESS}
	\end{align}

	After an observation is introduced into each model, we check the condition that the ESS is greater than some threshold, $E$, for example $J/2$. Once the ESS drops below this threshold, it indicates that the particles are less informative than a sample of $E$ independent draws from the target distribution. Using such a sample can lead to estimates of integrals with very high or even infinite variance.  Therefore, to tackle this problem and boost the ESS back up to $J$, we use a resampling algorithm.  Although this improves the value of the ESS, the sample will contain many duplicates. Therefore, after conducting this resampling step, a move step is required. 
	
	The purpose of a move step is to diversify the set of particles whilst maintaining invariance for the current target distribution. We do this by moving each particle according to a Markov Chain Monte Carlo (MCMC) kernel with invariant distribution $\pi_{i}(\vect{\theta}_{m, i}|m, \vect{y}_{1:i}, \vect{d}_{1:i})$. A disadvantage of using an MCMC kernel is that movement of all the particles is not guaranteed. Therefore, one iteration may not be enough to diversify the particle set. An appropriate number of times to conduct the move step was proposed by \citet{drovandi2011estimation} and must satisfy
	
	\begin{align}
	R_{m} \geq \frac{\log c}{\log (1-p)}.
	\label{eq:rt}
	\end{align}
	
	The value $1-c$ is our pre-specified probability that the particle will move and $p$ is the probability of acceptance at the MCMC move step. This acceptance probability, $p$, is estimated by conducting one ``probing" MCMC move step for each particle in the set and determining the overall proportion of particles which move.
	
	A useful property of this algorithm is that for each model $m$, we can approximate the log evidence, $\log{Z_{m,i}}$, using the  particle weights. \citet{del2006sequential} show that we can approximate the ratio of normalising constants, $Z_{m,i+1}/Z_{m,i}$, and hence the posterior predictive distribution, $f(y_{i+1}|m,\vect{y}_{1:i},\vect{d}_{1:i+1})$, for each model at the current experimental number $i$ using

	\begin{align*}
	Z_{m,i+1}/Z_{m,i} = f(y_{i+1}|m,\vect{y}_{1:i},\vect{d}_{1:i+1})
	&\approx  \sum_{j=1}^{J}W_{m,i}^{j }f(y_{i+1}|m,\vect{\theta}_{m,i}^{j}, d_{i+1}).
	\end{align*}	
	
	Since we know that $\log Z_{i+1} = \sum_{v=0}^{i} \log (Z_{i+1-v}/Z_{i-v})$ and $Z_{0} = 1$,  we are able to easily approximate $\log Z_{i+1}$ in our SMC algorithm. This can be achieved by adding the logarithm of the normalising constant ratio at each experimental number $i$ to the current log evidence: $\log Z_{m,i+1} = \log Z_{m,i} + \log (Z_{m,i+1}/Z_{m,i})$.
	
	Section \ref{sec:estimation} discusses a major benefit of using SMC for sequential experimental design, which is that SMC produces convenient outputs that are used for estimation of utilities and other important quantities.

	\subsection{Estimation of Utility Functions}
	\label{sec:estimation}
	SMC provides an efficient way for estimating the utility functions for proposed design points. The quantities within the utility functions are estimated solely using the particle values and the corresponding weights and are computed in the same way regardless of the model and parameters chosen for the data.  We now demonstrate how we can approximate $U(d,z,m|\vect{y}_{1:i}, \vect{d}_{1:i})$  and other relevant quantities such as posterior model probabilities appearing in algorithm \ref{alg:SMC}.   Define $w_{m,i}^{j}(d,z)$ and $W_{m,i}^{j}(d,z)$ to be the updated unnormalised and normalised weights of the $j^{th}$ particle after observing response $z$ at design $d$, respectively. We estimate the ratio of two normalising constants and hence the posterior predictive distribution by    
	
	\begin{align}
	\frac{Z_{m,i}(d,z)}{Z_{m,i}} = f(z|m,\vect{y}_{1:i},\vect{d}_{1:i},d) &\approx 
	\sum_{j=1}^{J} W_{m,i}^{j} f(z|\vect{\theta}_{m,i}^{j},d)= \sum_{j=1}^{J} w_{m,i}^{j}(d,z).
	\label{eq:evidence}
	\end{align}
	
	Using the normalised weighted samples and Monte Carlo integration, we can approximate the integral within the parameter estimation utility in \eqref{eq:util_pe2}: 
	
	\begin{align*}
	\int_{\vect{\theta}_{m}} \pi_{i+1}(\vect{\theta}_{m}|m, \vect{y}_{1:i},z,\vect{d}_{1:i}, d)~ \log{f(z|m, \vect{\theta}_{m},d)}~ \mathrm{d} \bm{\theta}_m
	\approx \sum_{j=1}^{J}W_{m,i}^{j}(d,z)\log{f(z|m, \vect{\theta}_{m,i}^{j}, d)}.
	\end{align*}
	
	We estimate the posterior predictive distribution,  ${f}(z|\vect{y}_{1:i}, \vect{d}_{1:i},d)$, by
	
	\begin{align*}
	{f}(z|\vect{y}_{1:i},\vect{d}_{1:i},d) \approx \sum_{m=1}^{K}\hat{\pi}_{i}(m|\vect{y}_{1:i},\vect{d}_{1:i})\sum_{j=1}^{J} w_{m,i}^{j}(d,z).
	\end{align*}
	
	The posterior model probabilities at experiment number $i$, ${\pi}_{i}(m|\vect{y}_{1:i},\vect{d}_{1:i})$, can be estimated by normalising the evidences (see \eqref{eq:evidence}). Using these approximations together with \eqref{eq:utility}, we can approximate the utilities. 	The SMC algorithm for optimal sequential experimental design is presented in algorithm \ref{alg:SMC}. 
	
	\begin{algorithm*}  
		\caption{SMC algorithm for sequential experimental design}
		
		\vspace{0.3cm}
		\textbf{INPUT:} Total number of experiments to run, $I$, number of samples for each model, $J$, an appropriate ESS threshold, $E$,  the model prior distributions, $\pi_{0}(\vect{\theta}_{m}|m)$, and the likelihood function, $f(y_{i}|m, \vect{\theta}_{m}, d_{i})$.\\
		
		\textbf{OUTPUT:} The selected design points, $\vect{d}_{1:I}$ and the responses observed at those design points, $\vect{y}_{1:I}$.\\
		\vspace{0.1cm}
		
		\begin{algorithmic}[1]
			\STATE Draw samples from model priors, $\vect{\theta}_{m,0}^{j} \sim \pi_{0}(\vect{\theta}_{m}|m)$, for $m = 1,...,K$ and for $j = 1,...,J$.
			\STATE Initialise weights, $W_{m,0}^{j} = 1/J$, for $m = 1,...,K$ and for $j = 1,...,J$.
			\STATE Initialise log evidences, $\log\hat{Z}_{m,0} = 0$, for $m = 1,...,K$.
			\FOR{$i = 0$ to $I-1$}
			\STATE Select design point $d_{i+1}$ to maximise some given utility $U(d|\vect{y}_{1:i}, \vect{d}_{1:i})$.
			\STATE Generate/collect observation $y_{i+1}$ at the design point $d_{i+1}$.
			\FOR{$m=1$ to $K$}
			\STATE Compute the updated unnormalised weights,  $w_{m,i+1}^{j} = W_{m,i}^{j}f(y_{i+1}|m, \vect{\theta}_{m,i}^{j}, d_{i+1})$, for $j=1,...,J$.
			\STATE Update the log evidence, $\log\hat{Z}_{m,i+1}= \log\hat{Z}_{m,i} + \log \sum_{j=1}^{J} w_{m,i+1}^{j}$.
			\STATE Normalise the weights, $W_{m,i+1}^{j} = w_{m,i+1}^{j} /\sum_{q=1}^{J} w_{m,i+1}^{q}$, for $j=1,...,J$.
			\STATE Compute the effective sample size, $\mbox{ESS}_{m, i+1} = 1 / \sum_{j=1}^{J}(W_{m,i+1}^{j})^{2}$ .
			\IF{$\mbox{ESS}_{m, i+1} < E$}
			\STATE Resample particle set to obtain $\left\{\vect{\theta}_{m,i+1}^{j}\right\}^{J}_{j=1}$.
			\STATE Set $W_{m,i+1}^{j} = 1/J$  for $j = 1,...,J$.
			\STATE Set $\mbox{ESS}_{m, i+1} = J$.
			\STATE Determine the parameters of the MCMC proposal $q_{m,i+1}(\cdot|\cdot)$ using the current particles, $\left\{\vect{\theta}_{m,i+1}^{j}\right\}^{J}_{j=1}$.
			\FOR{$j=1$ to $J$}
			\STATE Conduct a one iteration move step by moving the particle $\vect{\theta}_{m,i+1}^{j}$ with an MCMC kernel of invariant distribution  $\pi_{i+1}(\vect{\theta}_{m, i+1}|m, \vect{y}_{1:i+1}, \vect{d}_{1:i+1})$.
			\ENDFOR
			\STATE Calculate acceptance probability, $p$, and hence $R_{m}$.
			\FOR{$j=1$ to $J$}
			\STATE Move particle $\vect{\theta}_{m,i+1}^{j}$ with an MCMC kernel of invariant distribution $\pi_{i+1}(\vect{\theta}_{m,i+1}|m, \vect{y}_{1:i+1}, \vect{d}_{1:i+1})$ iterated $R_{m}-1$ times.
			\ENDFOR
			\ELSE
			\STATE Set $\vect{\theta}_{m, i+1}^{j} = \vect{\theta}_{m, i}^{j}$ for $j=1,...,J$.
			\ENDIF
			\ENDFOR
			\ENDFOR
		\end{algorithmic}
		\label{alg:SMC}
	\end{algorithm*}   
		
		\newpage
		
		\section{Simulation Study}
			\label{sec:simstudy}
	
	To demonstrate the methods highlighted in this paper, we now conduct a simulation study. The purpose of this simulation study is to illustrate the benefit of using optimal sequential design over optimal static design for predator-prey functional response experiments. For a detailed description of the optimal static design methodology used in this study, see Appendix A.  As a baseline to facilitate the comparison of optimal design methods, we also include a design strategy that randomly selects initial prey levels. 
	
	For our simulation study, we consider 40 different settings for the true data generating process, consisting of 4 different true models each with 10 different true parameter settings. The four different models are defined in Table 1.\\

	\begin{center}
		
		\begin{tabular}{c c|c|}
			\cline{2-3}
			&\multicolumn{2}{|c|}{\textbf{Mechanistic Model}} \\ 
			\hline
			\multicolumn{1}{|c|} {\textbf{Distribution of \vect{$n$}}} 
			& \multicolumn{1}{c|} {Holling\textquotesingle s type II} 
			& \multicolumn{1}{c|} {Holling\textquotesingle s type III}\\
			\hline
			\multicolumn{1}{|c|} {beta-binomial} 
			& \multicolumn{1}{c|} {Model 1}
			& \multicolumn{1}{c|} {Model 2} \\ 
			\multicolumn{1}{|c|} {binomial} 
			& \multicolumn{1}{c|} {Model 3 }
			& \multicolumn{1}{c|} {Model 4} \\ 
			\hline
			\label{tab:models}	
		\end{tabular}
		
		Table 1: Models used for simulation study\\	
	\end{center}
	
For each of the 40 different model/parameter configurations, and for each design selection method, we run algorithm \ref{alg:SMC} 30 times independently. The total number of observations collected in each run of the SMC algorithm is $I = 25$. The set of true parameter settings are drawn from the posterior distribution based on the dataset from Papanikolaou et al. (2016) (we use SMC to sample the posterior here). The dataset used is the same data from Figure \ref{fig:example}.

For this simulation study, we consider the total exposure time of prey and predator to be $\tau = 24$ hours. The design points $N_{0,i}$ are members of the set $D = \{1,\ldots,300\}$ and the responses can assume the values $n_{i} \in \{0,\ldots,N_{0,i}\}$ for $i = 1,\ldots,I$. The prior distribution of the parameters is given by $\log(a) \sim \mbox{N}(-1.4, 1.35^{2})$, $\log(T_{h}) \sim \mbox{N}(-1.4, 1.35^{2})$ and $\log(\lambda) \sim \mbox{N}(-1.4, 1.35^{2})$. The prior distribution was chosen to be reasonably vague but giving little support to unrealistic regions of the parameter space.  If the practitioner has access to more prior information, this can easily be incorporated into our framework by simply changing the prior distribution. The parameter $\lambda$ is only required in the beta-binomial models. In addition, the prior model probabilities are equal across the $K=4$ models.

To assess the effect of the different design point selection methods on the precision of the parameters, we compare the distribution of the Bayesian D-posterior precision \citep{drovandi2013sequential} obtained by the simulation study across the different methods. The Bayesian D-posterior precision, which is estimated by taking the inverse of the determinant of the weighted sample covariance matrix of the SMC particles, is a measure that allows us to quantify the precision of model parameters. The comparison of the Bayesian D-posterior precision is conducted for each true model. Similarly, to assess the model discrimination power of the algorithm, we can compare the distribution of posterior model probabilities for each model across the different design methods and different true models.  Section \ref{sec:results} explores these comparisons and displays results from the simulation study.

	\subsection{Results}
	\label{sec:results}
	
\begin{figure}[!htp]
		\centering
		\subfigure[True Model 1]{\includegraphics[height=0.18\textheight,keepaspectratio]{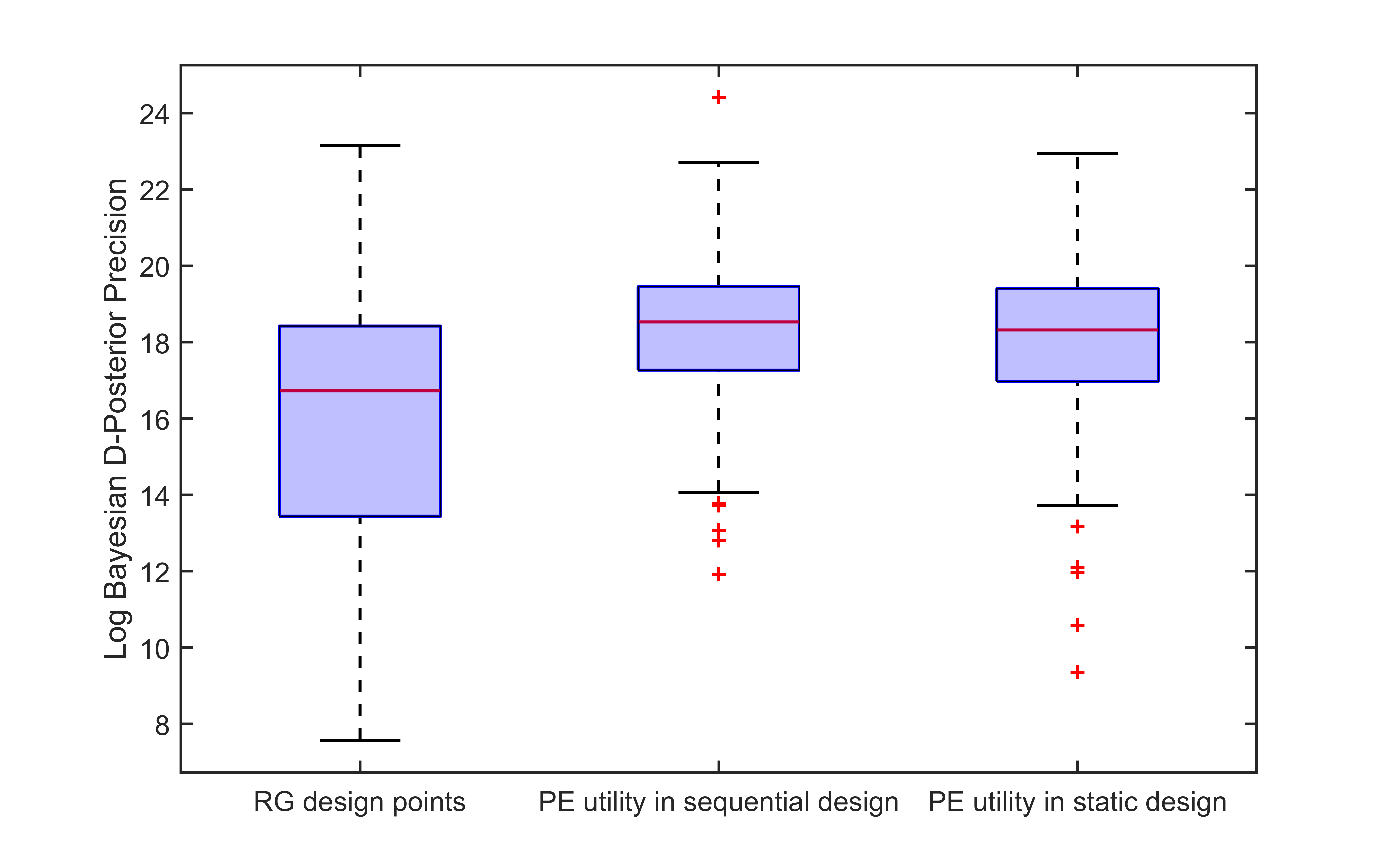}\label{figsub:A_pea}}
		\subfigure[True Model 2]{\includegraphics[height=0.18\textheight,keepaspectratio]{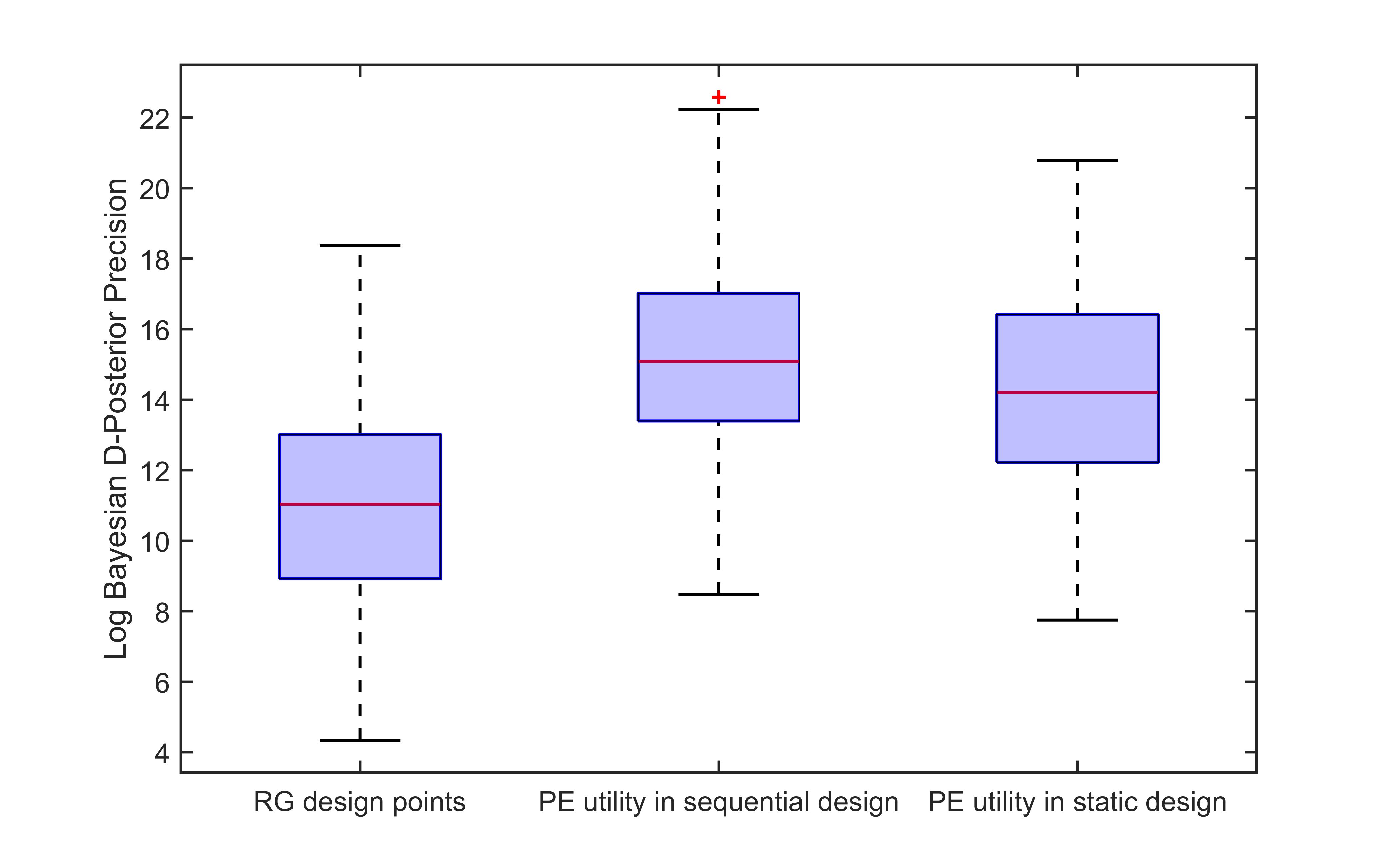}\label{figsub:A_peb}}
		\subfigure[True Model 3]{\includegraphics[height=0.18\textheight,keepaspectratio]{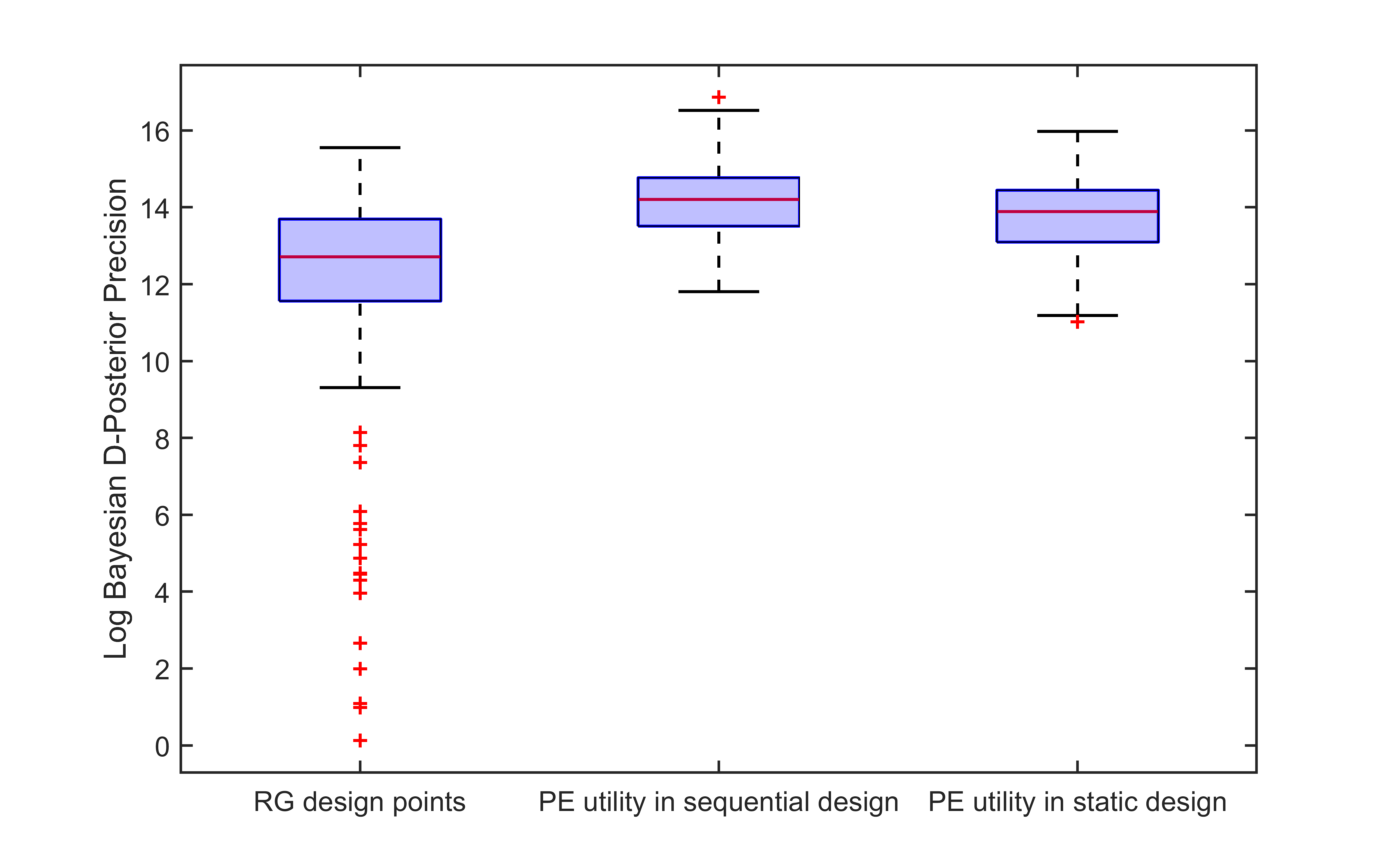}\label{figsub:A_pec}}
		\subfigure[True Model 4]{
			\includegraphics[height=0.18\textheight,keepaspectratio]{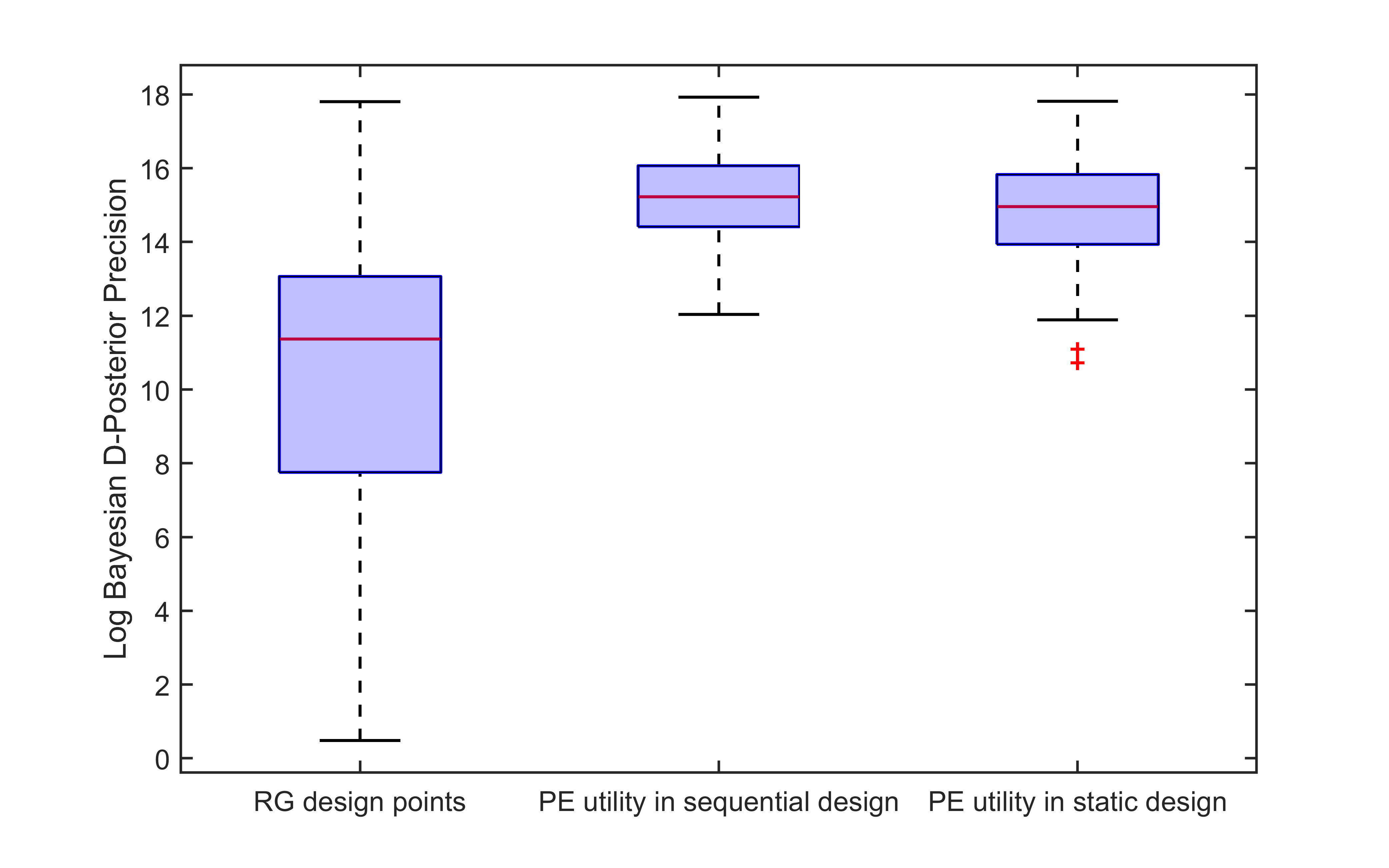}\label{figsub:A_ped}}
		\caption{Distribution of the log Bayesian D-posterior precision for random, sequential and static design selection methods for each true model. RG and PE signify randomly generated and parameter estimation, respectively. The goal of the optimal designs is precise parameter estimation.}
		\label{fig:A_pe}
	\end{figure}

	For each true model, Figure \ref{fig:A_pe} compares the distribution of the log Bayesian D-posterior precision for the designs generated from the random, sequential and static methodologies. In this particular case, the experimental aim for the static and sequential designs is precise parameter estimation.	It is apparent from Figure \ref{fig:A_pe} that the sequential experimental design methodology outperforms the other methods for all true models. In addition, we can see that the two optimal designs outperform the randomly generated designs. Therefore, both optimal experimental design methods can be beneficial for parameter estimation in predator-prey functional response experiments. 
	
\begin{figure}[!htp]
		\centering
		\subfigure[True Model 1]{\includegraphics[	height=0.185\textheight,keepaspectratio]{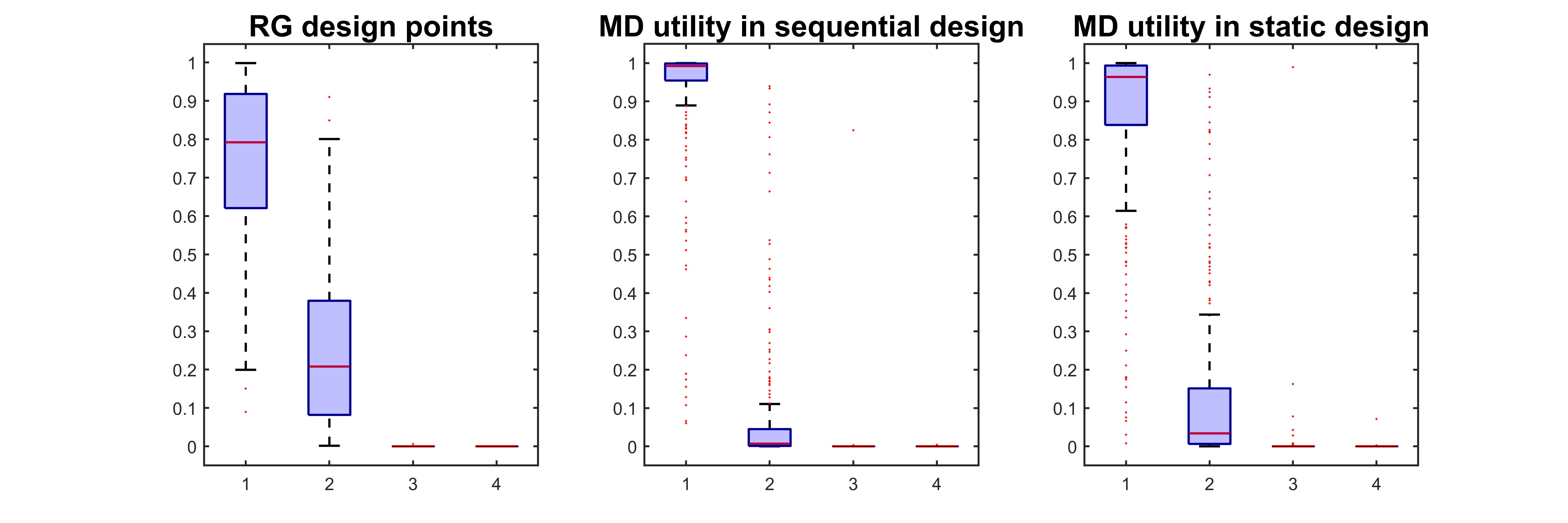}\label{figsub:A_mda}}
		\subfigure[True Model 2]{\includegraphics[	height=0.185\textheight,keepaspectratio]{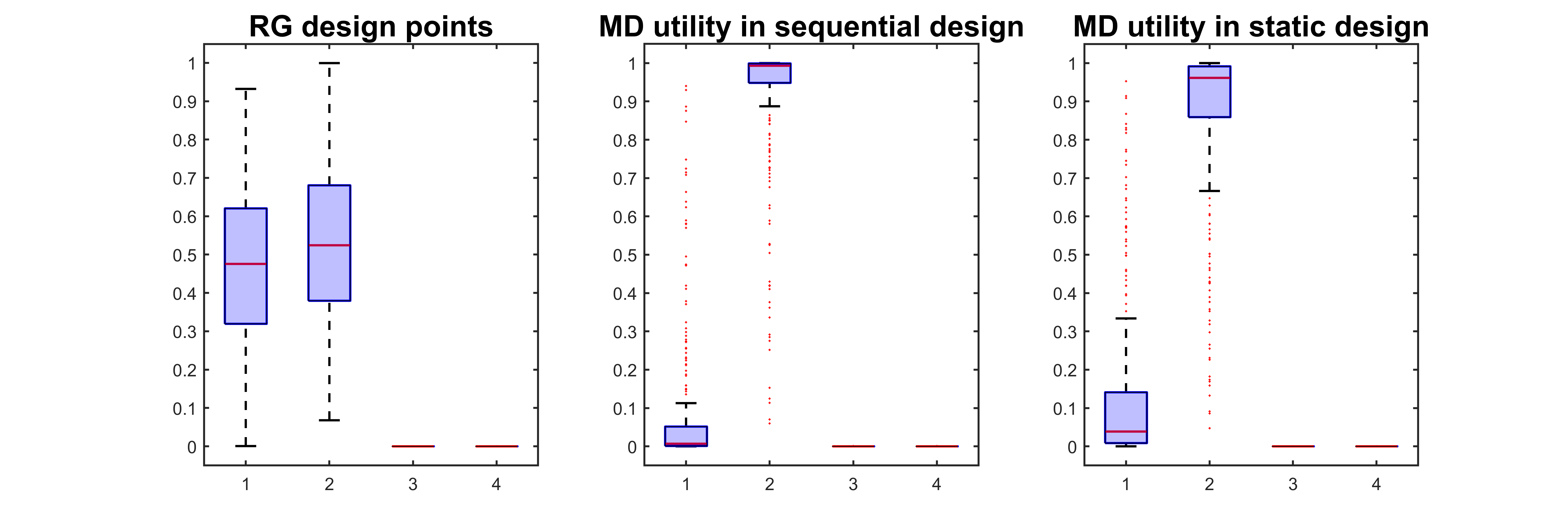}\label{figsub:A_mdb}}
		\subfigure[True Model 3]{\includegraphics[	height=0.185\textheight,keepaspectratio]{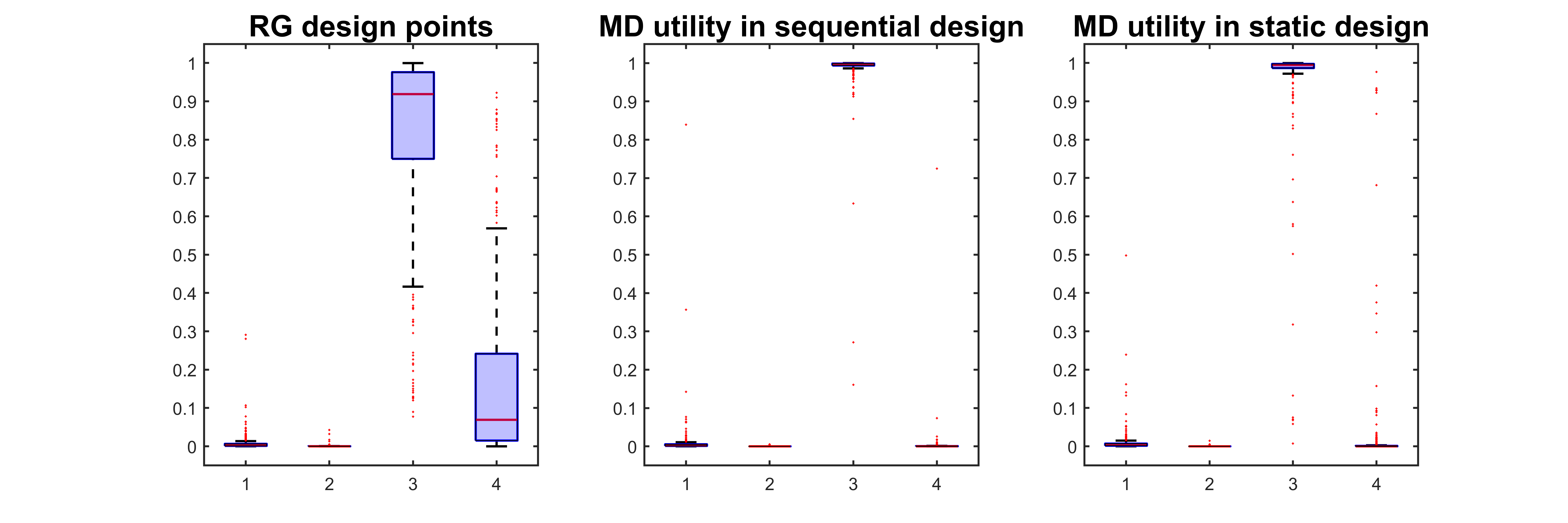}\label{figsub:A_mdc}}
		\subfigure[True Model 4]{\includegraphics[	height=0.185\textheight,keepaspectratio]{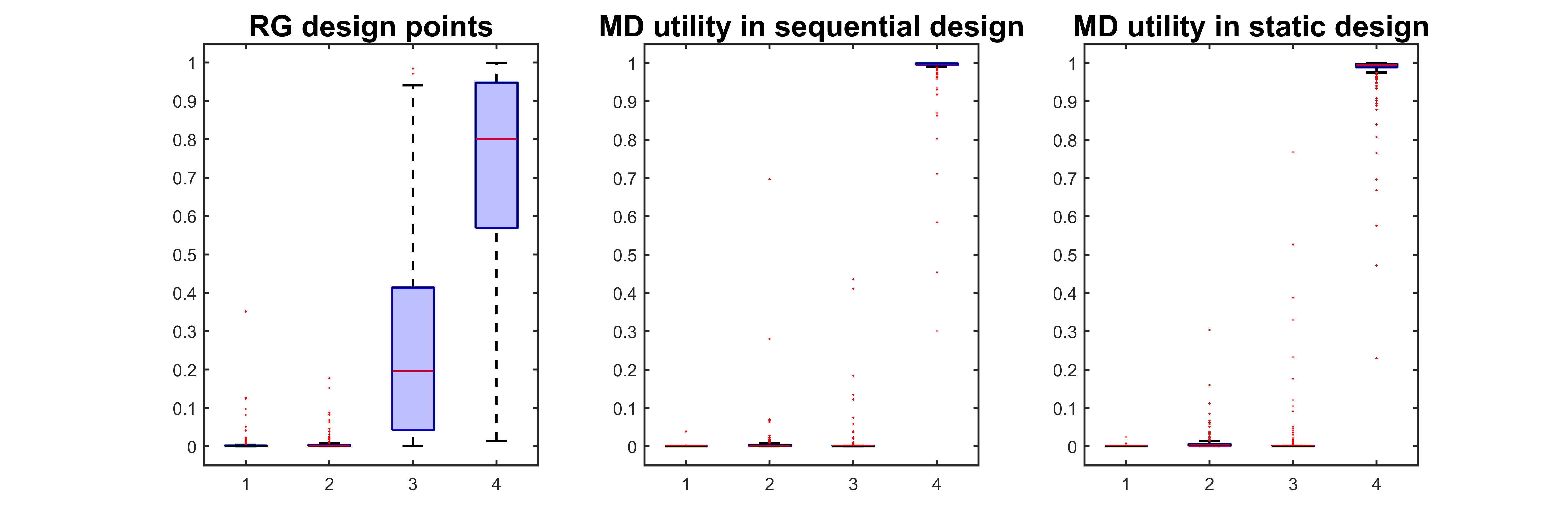}\label{figsub:A_mdd}}
		\caption{Distributions of the posterior model probabilities for randomly generated, sequential and static design selection methods. Distributions are displayed and compared for each true model. RG and MD signify randomly generated and model discrimination, respectively. The x-axis on each plot represents the model number and corresponds to the numbers in Table 1. The y-axis represents the posterior model probability. The goal of the optimal designs is discriminating between models.}
		\label{fig:A_md}
	\end{figure}

	The distributions of the posterior model probabilities obtained with randomly generated, optimal sequential and optimal static designs for different true models are shown in Figure \ref{fig:A_md}. The goal of the optimal designs in this example is to discriminate between models. Similar to the results of the parameter estimation experimental goal, the sequential design outperforms the other methods for the goal of model discrimination.

\begin{figure}[!htp]
		\centering
		\subfigure[True Model 1]{\includegraphics[height=0.18\textheight,keepaspectratio]{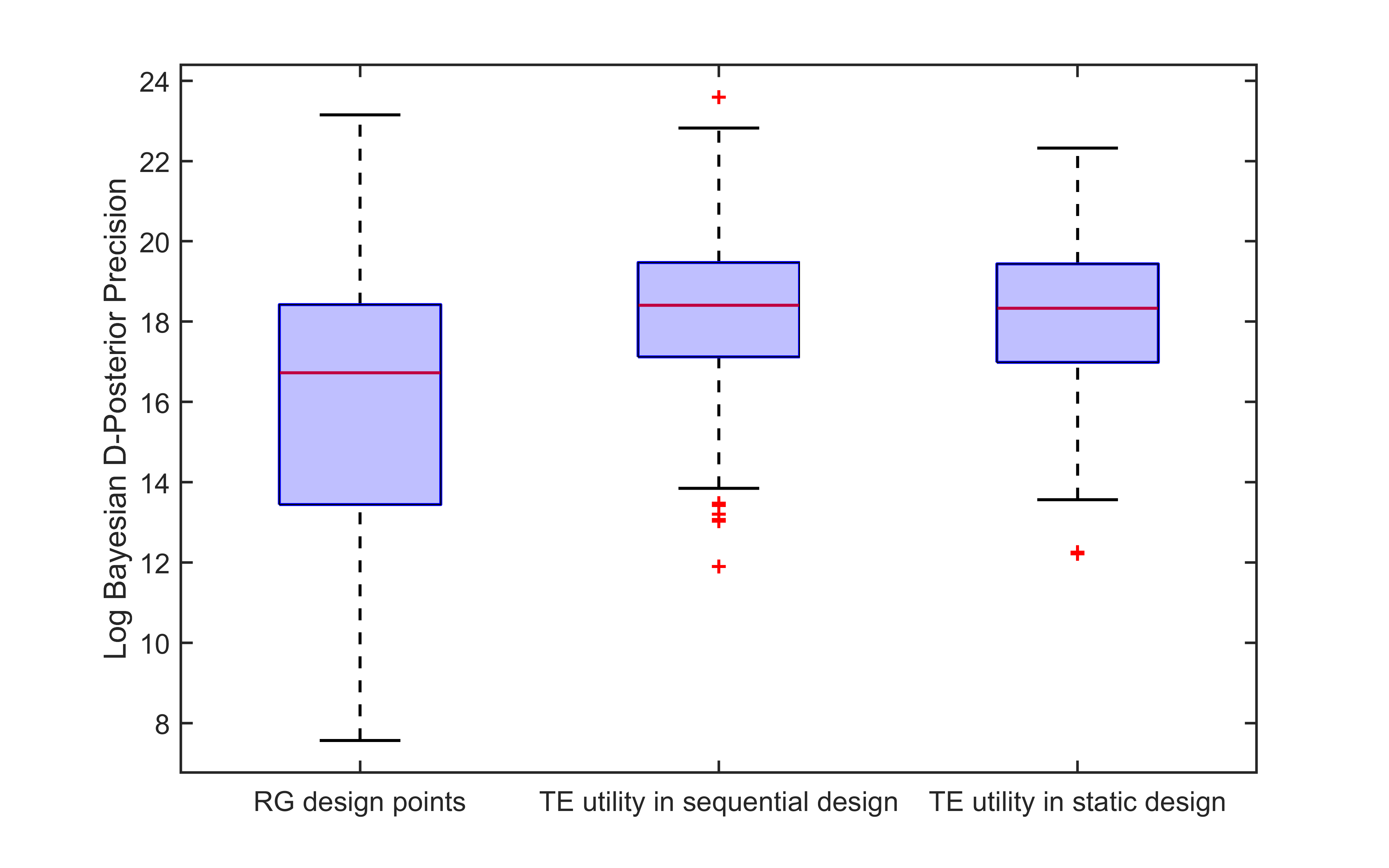}\label{figsub:A_pe2a}}
		\subfigure[True Model 2]{\includegraphics[height=0.18\textheight,keepaspectratio]{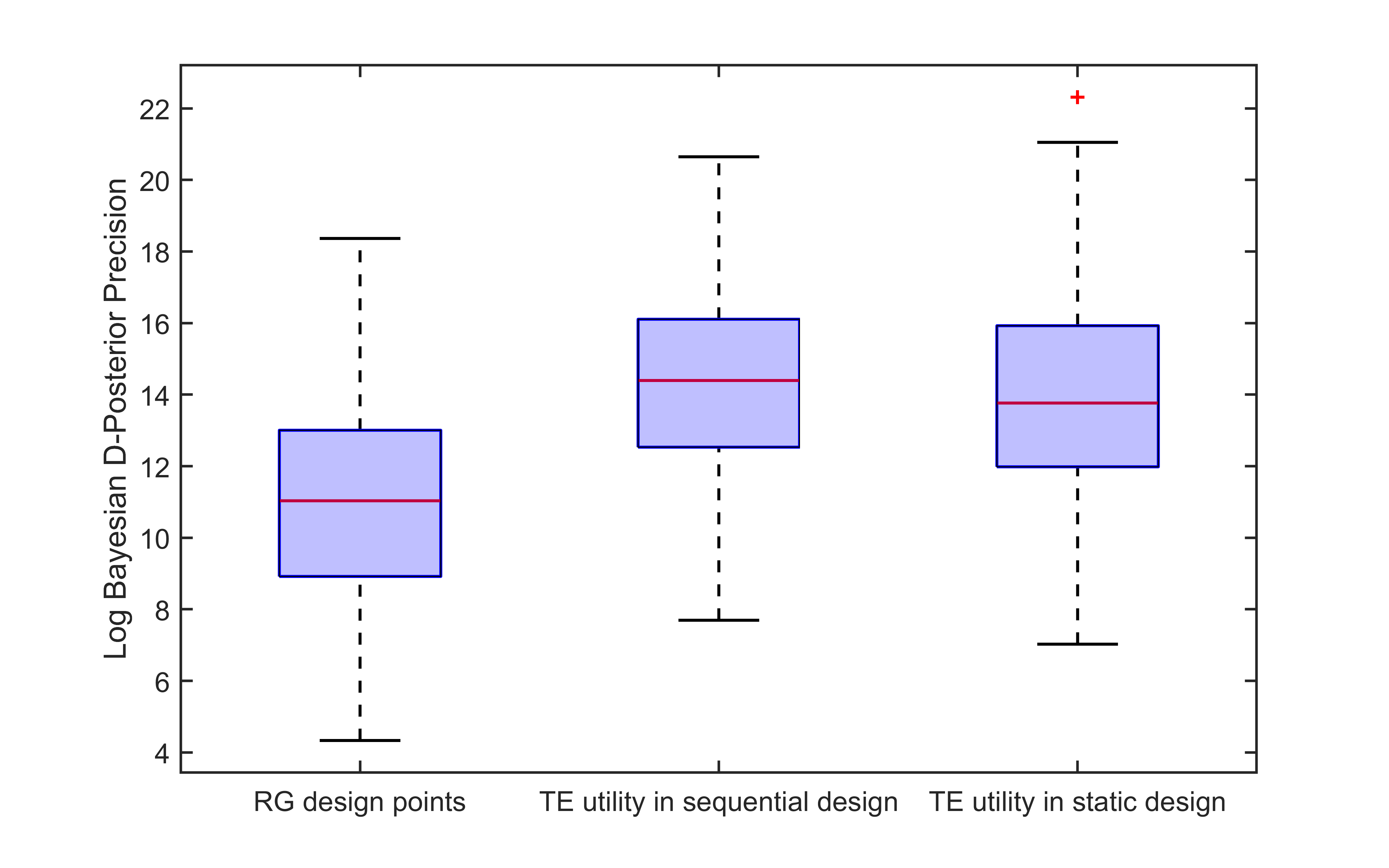}\label{figsub:A_pe2b}}
		\subfigure[True Model 3]{\includegraphics[height=0.18\textheight,keepaspectratio]{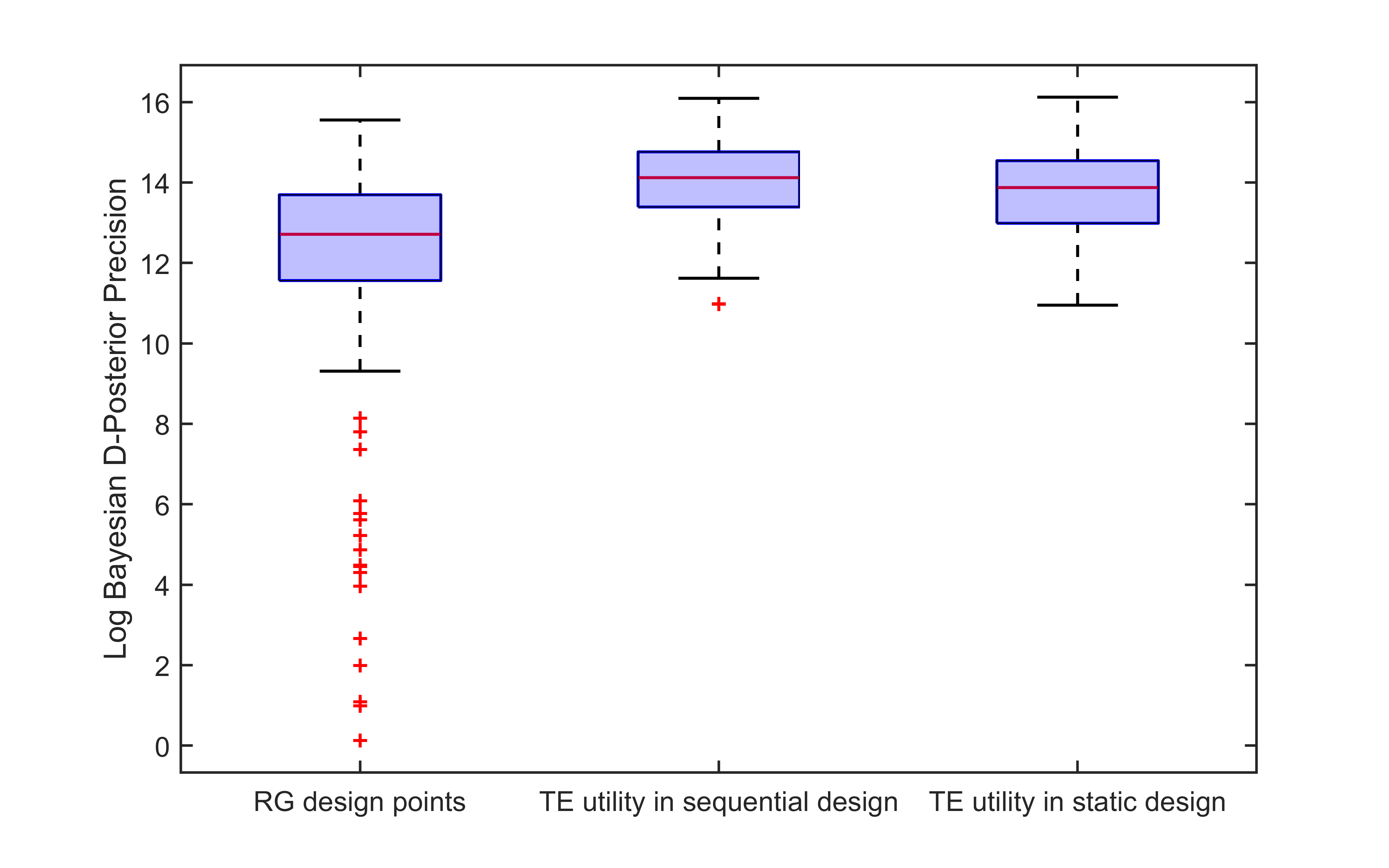}\label{figsub:A_pe2c}}
		\subfigure[True Model 4]{
			\includegraphics[height=0.18\textheight,keepaspectratio]{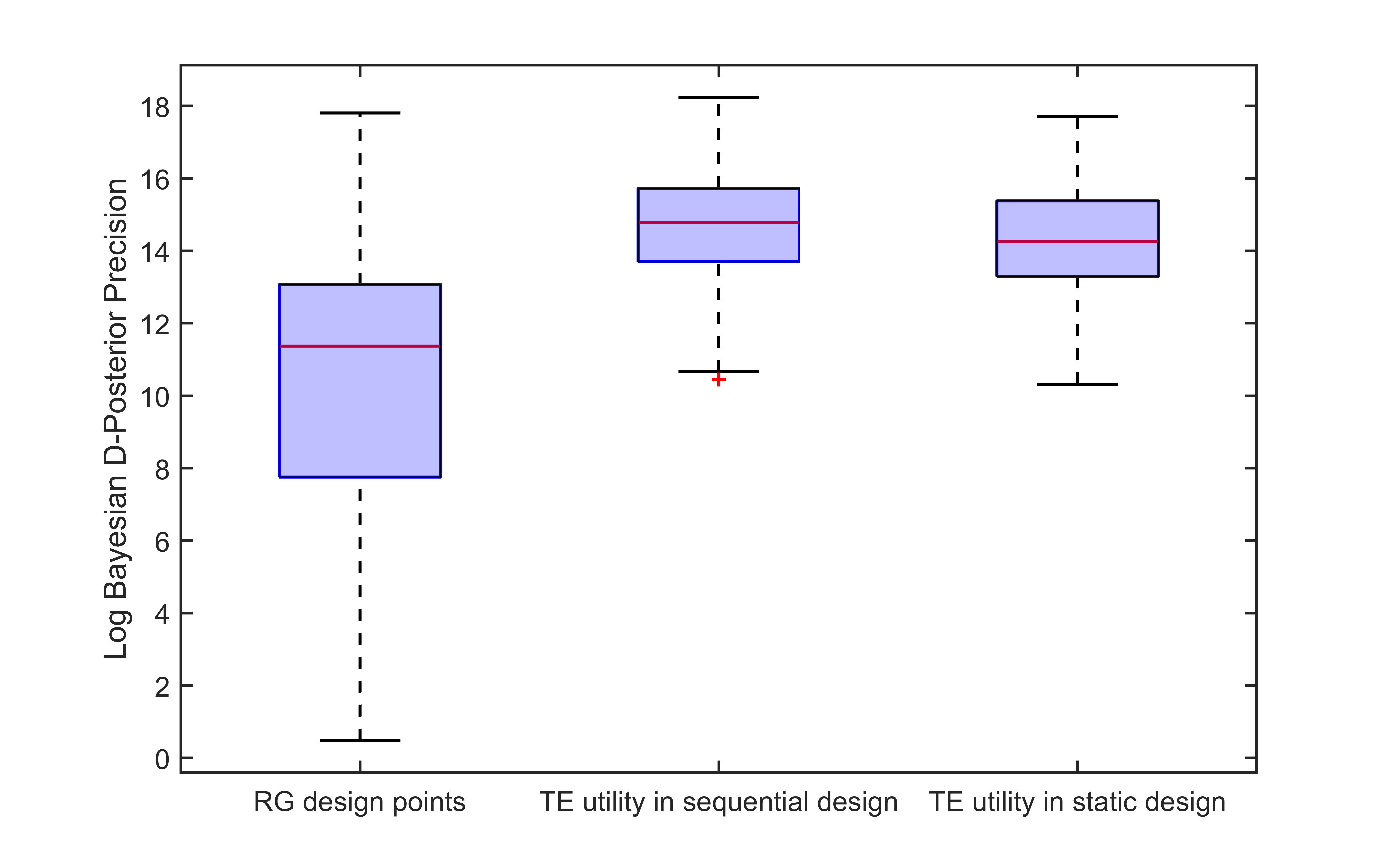}\label{figsub:A_pe2d}}
		\caption{Distributions of the log Bayesian D-posterior precision for random, sequential and static design selection methods for each true model. RG and TE signify randomly generated and total entropy, respectively. The goal of the optimal designs is dual-purpose (parameter estimation and model discrimination).}
		\label{fig:A_pe2}
	\end{figure}

\begin{figure}[!htp]
		\centering
		\subfigure[True Model 1]{\includegraphics[	height=0.19\textheight,keepaspectratio]{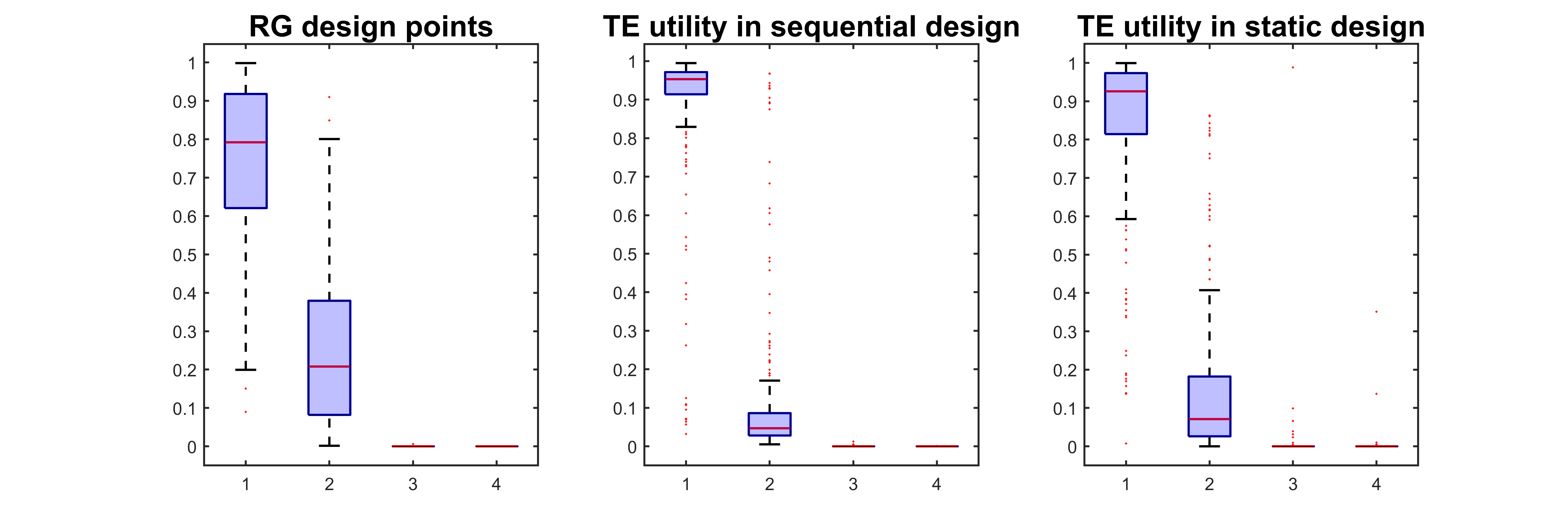}\label{figsub:A_md2a}}
		\subfigure[True Model 2]{\includegraphics[	height=0.19\textheight,keepaspectratio]{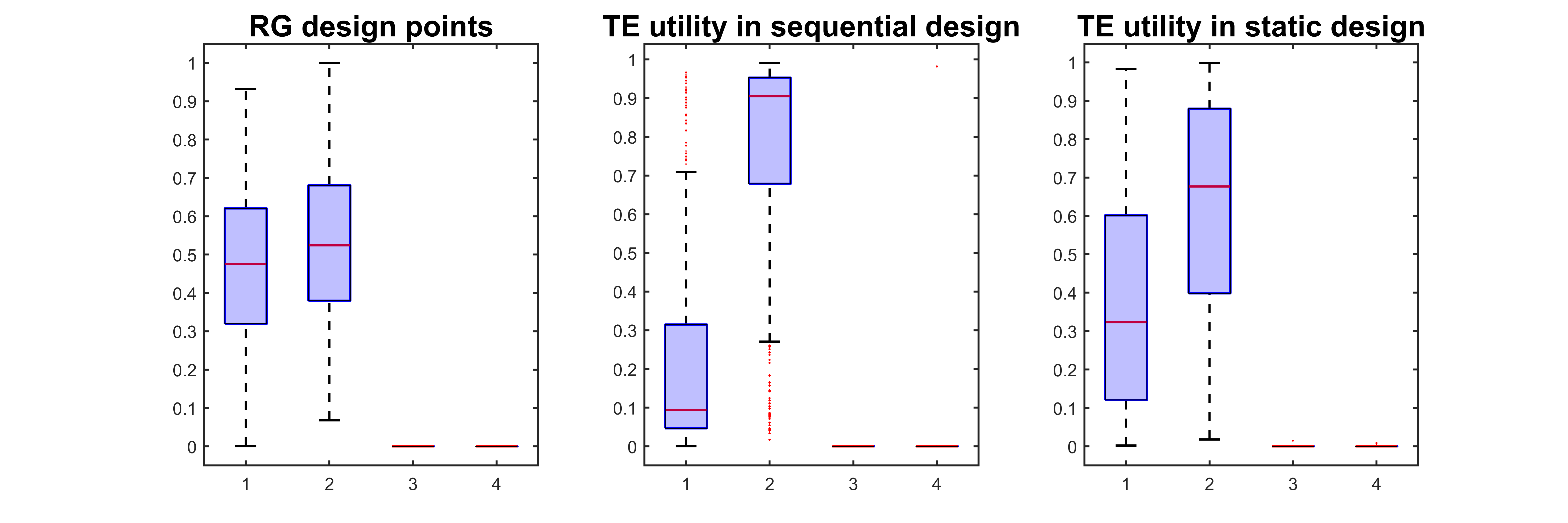}\label{figsub:A_md2b}}
		\subfigure[True Model 3]{\includegraphics[	height=0.19\textheight,keepaspectratio]{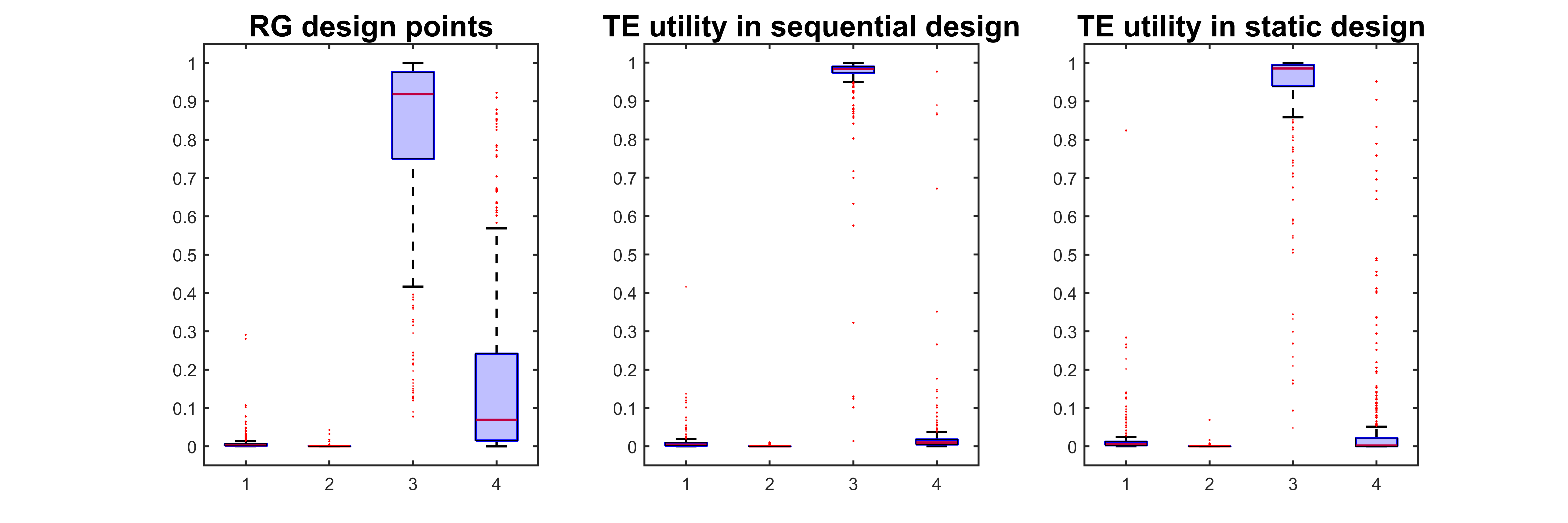}\label{figsub:A_md2c}}
		\subfigure[True Model 4]{
			\includegraphics[	height=0.19\textheight,keepaspectratio]{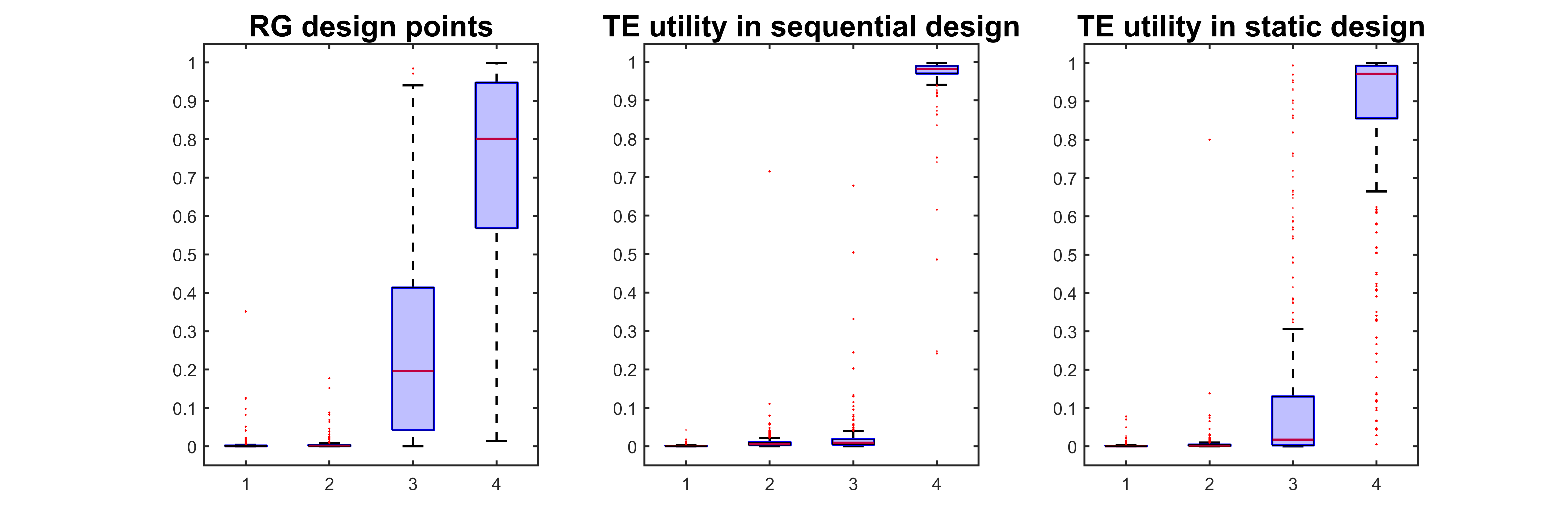}\label{figsub:A_md2d}}
		\caption{Distributions of the posterior model probabilities for randomly generated, sequential and static design selection methods. Distributions are displayed and compared for each true model. RG and TE signify randomly generated and total entropy, respectively. The x-axis on each plot represents the model number and corresponds to the numbers in Table 1. The y-axis represents the posterior model probability. The goal of the optimal designs is dual-purpose (parameter estimation and model discrimination).}
		\label{fig:A_md2}
	\end{figure}

	Figure \ref{fig:A_pe2} and Figure \ref{fig:A_md2} compare the precision of parameters and model probabilities, respectively, between randomly generated, sequential and static designs when the experimental aim is dual-purpose. We can  see from the figures that the sequential design outperforms the static for both components of the dual-purpose experimental goal. This pattern is consistent across all true models.

	\section{Discussion}
	\label{sec:diss}
	
	Optimally designed predator-prey functional response experiments are largely advantageous for quantitative ecological studies.  Optimal designs in functional response can reduce cost, improve time efficiency and prevent the sacrificing of animals. Previously, the only approach for optimal experimental design within predator-prey functional experiments \citep{zhang2018optimal} has been a static design which is solely reliant on the information known prior to the experiment. Therefore, current methodology does not take into account the information collected during the experiment. This can lead to inefficient designs and outcomes which are not significantly informative. Using sequential designs over static designs enables practitioners to update their information on model and parameter uncertainty as observations are collected sequentially. 
	
	This paper outlines the first sequential experimental design method for predator-prey functional response experiments. It utilises SMC, which is computationally efficient and allows the convenient estimation of utility functions for parameter estimation or model discrimination.  The results from our simulation study highlight the advantage of using optimal sequential design over other alternatives such as optimal static design and randomly generated designs. Additionally, the results in Appendix B illustrate that the dual-purpose design selection method is able to effectively discriminate between functional response models and acquire precise parameter estimation simultaneously. 
	
	Computational issues with our algorithm occasionally arise when fitting the incorrect model to the data. After a new observation is collected, the particle weights for each model are updated in our SMC algorithm. If we are fitting the incorrect model to the data, there can be a large difference between the posterior at experiment number $i$ and the posterior at experiment number $i+1$. This results in the ESS at $i+1$ becoming extremely low, even to the point where the ESS is approximately 1. If this occurs,  only a few unique particles remain in the sample, even after the resampling and move steps are completed. Using this sample can lead to poor estimates of utility functions and other quantities. These computational issues could possibly be avoided by constructing a sequence of target distributions between the posterior at experiment number $i$ and at experiment number $i+1$.
	
	Another obvious limitation to this myopic approach is that looking only one observation ahead is not optimal \citep{borth1975total}. We could possibly  investigate a two-step ahead dynamic programming algorithm and determine whether it has a significant effect on results. 
	
	The ability to use optimal design to help achieve the experimental goals of parameter estimation and model discrimination is highly desirable in any application. The conducted simulation study illustrates that our sequential experimental design algorithm is a useful tool for predator-prey functional response experiments. Our algorithm enables a conclusion regarding model and parameter uncertainty to be reached at a much earlier stage than would be possible with a random or static design.
	
	\section*{Data Accessibility}
	R code for implementing the methodology is available via \url{https://github.com/haydenmoffat/sequential_design_for_predator_prey_experiments}.
	
	\section*{Authors' Contributions}
	C.D. designed the study. H.M., M.H. and C.D. developed the methods.  H.M. wrote the computer code, performed the analysis, interpreted results and drafted the paper. N.E.P. provided ecological expertise.  C.D., M.H. and N.E.P critically reviewed the paper.
	
	\section*{Competing Interests}
	We declare that we have no competing interests.
	
	\section*{Funding}
	HM was supported by a QUT Master of Philosophy Award. MH was funded by the Austrian Science Fund (FWF): J3959-N32. NEP gratefully acknowledges financial support from the Foundation for Education and European Culture under a postdoctoral research grant. CD was supported by an Australian Research Council Discovery Project (DP200102101).
	
	\section*{Acknowledgements}
	
	The authors would like to thank Dr. Theodore Kypraios of the University of Nottingham for providing feedback on an earlier draft. Computational resources and services used in this work were provided by the HPC and Research Support Group, Queensland University of Technology, Brisbane, Australia.  The authors thank the two anonymous reviewers for providing thoughtful comments that have helped improve the manuscript.
	
	\bibliographystyle{apalike} 
	\bibliography{References}

	\newpage
	
	\section*{Appendix A}
	
	For completeness, we have included a description of the methodology used to conduct optimal static design. We consider optimal designs for the purpose of precise parameter estimation and/or model discrimination. 
	
	\subsection*{Static Optimal Design Methodology}
	
	Bayesian optimal static design involves using prior information to select multi-dimensional designs for a particular experimental goal. Obtaining an optimal design in this way is computationally challenging as it requires maximising an analytically intractable utility function over a high-dimensional design space. To determine these designs, we use the methodology of \citet{overstall2018approach} to find a multi-dimensional near-optimal static design. 
	
	\subsubsection*{General Methodology}
	
	Recall the notation used within the optimal design section of this paper. For a given experiment, the total number of observations is $I$. The user-specified utility function for the model $m$, the design $\vect{d}_{1:I}$ and the response $\vect{y}_{1:I}$ is denoted $U(\vect{d}_{1:I}, \vect{y}_{1:I}, m)$. Here we do not explicitly include $\vect{\theta}_{m}$ in the utility function because the utilities that we define in the next subsection and in the main paper have $\vect{\theta}_{m}$ integrated out.	
	
	The optimal static design is found by maximising the expected utility over the design space. The expectation is taken with respect to the joint distribution of all unknown quantities, which include the models and experimental responses. The expectation of the user-specified utility, and thus the utility of the multi-dimensional design, $\vect{d}_{1:I}$, is given by
	
	\begin{align*}
	U(\vect{d}_{1:I}) = E_{m, \vect{y}_{1:I}| \vect{d}_{1:I}} \left[U(\vect{d}_{1:I}, \vect{y}_{1:I}, m) \right].	
	\end{align*}
	
	Since the expectation of the utility function typically does not have a closed-form solution,  numerical methods such as Monte Carlo estimation are required to approximate it. The Monte Carlo approximation of the expected utility is given by
	
	\begin{align*}
	U(\vect{d}_{1:I}) = \frac{1}{B} \sum_{b=1}^{B} U(\vect{d}_{1:I},\vect{y}_{1:I}^{b}, m^{b} ),
	\end{align*}
	
	where $B$ represents the number of Monte Carlo draws and $\left\lbrace \vect{y}_{1:I}^{b}, m^{b}  \right\rbrace_{b = 1}^{B} $ are samples from the joint distribution of $m \mbox{ and } \vect{y}_{1:I}$. Therefore, to compute this approximation for the expected utility, we only require the ability to sample from the joint distribution of $m \mbox{ and } \vect{y}_{1:I}$ and to evaluate the user-specified utility function for each sample. The evaluation of the utility functions is the focus of the next subsection.
	
	\newpage
	
	\subsubsection*{Utility Functions}
	
	The following utility functions can be used to encapsulate the goals of the experiment (e.g. parameter estimation, model discrimination and dual-purpose). The parameter estimation utility for model $m$ is given by the KLD between the prior distribution of $\vect{\theta}_{m}$, $\pi_{0}(\vect{\theta}_{m}|m)$, and the posterior distribution of $\vect{\theta}_{m}$, $\pi_{I}(\vect{\theta}_{m}|m,\vect{y}_{1:I},\vect{d}_{1:I})$. The parameter estimation utility is given by
	
	\begin{align}
	U(\vect{d}_{1:I}, \vect{y}_{1:I}, m)= \int_{\vect{\theta}_{m}} \pi_{I}(\vect{\theta}_{m}|m,\vect{y}_{1:I},\vect{d}_{1:I}) ~ \log{\left(\frac{ \pi_{I}(\vect{\theta}_{m}|m,\vect{y}_{1:I},\vect{d}_{1:I})}{\pi_{0}(\vect{\theta}_{m}|m)}\right)} ~ \mathrm{d}\vect{\theta}_{m}.
	\label{eq:A_utilpe}
	\end{align}
	
	The static design utility in \eqref{eq:A_utilpe} differs from the sequential parameter estimation utility function given in the main text in \eqref{eq:util_pe2}. In sequential design, data is collected one observation at a time. Therefore, utilities in these sequential designs are based on a singular design point rather than a mutli-dimensional design. Consequently, the sequential parameter estimation utility is given by the KLD between the current (after $i$ observations are collected) and updated (after $i+1$ observations are collected) posterior  distributions rather than the prior (before any data is collected) and posterior (after all data is collected) distributions of the parameters. 
	
	The \citet{box1967discrimination} utility for model discrimination is given by
	
	\begin{align}
	U(\vect{d}_{1:I}, \vect{y}_{1:I}, m) =  \log \pi_{I}(m|\vect{y}_{1:I},\vect{d}_{1:I}).
	\label{eq:A_utilmd}
	\end{align}
	
	The dual-purpose utility suggested by \citet{borth1975total} is given by the sum of the parameter estimation and model discrimination utilities. In all three utilities, posterior quantities are required to evaluate the utility function. Since the posterior distribution of $\vect{\theta}_{m}$ and $m$ will not have a neat closed form, a further approximation is required.

	\subsubsection*{Approximating Posterior Distributions and Other Quantities}
	
	To make inferences on models/parameters in the Bayesian framework, the joint posterior distribution of the parameters $\vect{\theta}_{m}$ and model $m$ is of interest. This joint distribution can be represented in a convenient form: $\pi_{I}(\vect{\theta}_{m}, m |\vect{y}_{1:I}, \vect{d}_{1:I}) = \pi_{I}(\vect{\theta}_{m}|m,\vect{y}_{1:I}, \vect{d}_{1:I})\pi_{I}(m|\vect{y}_{1:I}, \vect{d}_{1:I})$. For more information on Bayesian inference in the presence of parameter and model uncertainty, see \citet{o2004kendall}. The posterior distribution of parameters is given by
	
	\begin{align}
	\pi_{I}(\vect{\theta}_{m}|m, \vect{y}_{1:I}, \vect{d}_{1:I}) = \frac{f(\vect{y}_{1:I}|m,\vect{\theta}_{m}, \vect{d}_{1:I})\pi_{0}(\vect{\theta}_{m}|m)}{f(\vect{y}_{1:I}|m,\vect{d}_{1:I})},
	\end{align}
	
	where the marginal likelihood is given by
	
	\begin{align*}
	f(\vect{y}_{1:I}|m,\vect{d}_{1:I}) = \int_{\vect{\theta}_{m}} f(\vect{y}_{1:I}|m,\vect{\theta}_{m}, \vect{d}_{1:I})\pi_{0}(\vect{\theta}_{m}|m)~ \mathrm{d}\vect{\theta}_{m}.
	\end{align*}
	
	\newpage
	
	The posterior model probabilities are given by 
	
	\begin{align}
	\pi_{I}(m|\vect{y}_{1:I}, \vect{d}_{1:I}) = \frac{f(\vect{y}_{1:I}|m, \vect{d}_{1:I})\pi_{0}(m)}{\sum_{k=1}^{K}f(\vect{y}_{1:I}|k, \vect{d}_{1:I})\pi_{0}(k)},
	\label{eq:A_mp}
	\end{align}
	
	where the $k$ inside the summation represents $M=k$. As displayed in \eqref{eq:A_mp}, to compute the posterior model probability for $ m ~ \epsilon~ \left\lbrace 1,...,K \right\rbrace$, we require the marginal likelihood, $f(\vect{y}_{1:I}|m,\vect{d}_{1:I})$, which can be analytically intractable. \citet{overstall2018approach} proposed the use of normal-based Laplace approximations to approximate distributions of interest for static Bayesian designs. Using this approach, the marginal likelihood can be constructed by using only the posterior mode and the Fisher information. The posterior mode for model $m$, $\hat{\vect{\theta}}_{m}(\vect{y}_{1:I})$, can be computed by $\hat{\vect{\theta}}_{m}(\vect{y}_{1:I}) = \mbox{arg }  \underset{\vect{\theta}_{m}}{\mbox{max }} f(\vect{y}_{1:I}|m,\vect{\theta}_{m}, \vect{d}_{1:I})\pi_{0}(\vect{\theta}_{m}|m)$. Let $\mathcal{I}(\vect{\theta}_{m}|m)$ denote the Fisher information for $\vect{\theta}_{m}$ given $m$ and $\mathbf{H}(\vect{\theta}_{m};m)$ denote the Hessian of the negative log joint density of $\vect{\theta}_{m}$ and $m$. The covariance of the Laplace approximation is given by 
	
	\begin{align*}
	\hat{\vect{\Sigma}}_{m}(\vect{y}_{1:I}) = \mathbf{H}(\hat{\vect{\theta}}_{m}(\vect{y}_{1:I});m)^{-1},
	\end{align*}
	
	where
	
	\begin{align*}
	\mathbf{H}(\vect{\theta}_{m};m) & =\mathcal{I}(\vect{\theta}_{m}|m)- \frac{\partial^{2} \log \pi_{0}(\vect{\theta}_{m}| m)}{\partial \vect{\theta}_{m}\vect{\theta}_{m}^{T}}\\
	&= \mathrm{E}_{\vect{y}_{1:I}|m,\vect{\theta}_{m}}\left[\frac{\partial \log f(\vect{y}_{1:I}|m,\vect{\theta}_{m}, \vect{d}_{1:I})}{\partial \vect{\theta}_{m}}\frac{\partial \log f(\vect{y}_{1:I}|m,\vect{\theta}_{m}, \vect{d}_{1:I})}{\partial \vect{\theta}_{m}^{T}} \right] - \frac{\partial^{2} \log \pi_{0}(\vect{\theta}_{m}| m)}{\partial \vect{\theta}_{m}\vect{\theta}_{m}^{T}}.
	\end{align*}
	
	The Laplace approximation to the marginal likelihood is given by
	
	\begin{align*}
	\tilde{f}(\vect{y}_{1:I}|m,\vect{d}_{1:I}) = (2\pi)^{\frac{p_{m}}{2}}|\hat{\vect{\Sigma}}_{m}(\vect{y}_{1:I})|^{\frac{1}{2}} f(\vect{y}_{1:I}|\hat{\vect{\theta}}_{m}(\vect{y}_{1:I}), m, \vect{d}_{1:I} )\pi_{0}(\hat{\vect{\theta}}_{m}(\vect{y}_{1:I})| m),
	\end{align*}
	
	where $p_{m}$ is the number of parameters in model $m$. In practice, we can estimate the relevant quantities required for the Laplace approximation (i.e. the posterior mode, Hessian) using a numerical optimiser. The estimate of the Hessian matrix is determined using finite differencing. Using the Laplace approximation of the marginal likelihood, the posterior model probabilities can be approximated by
	
	\begin{align*}
	\tilde{\pi}(m|\vect{y}_{1:I}, \vect{d}_{1:I}) = \frac{\tilde{f}(\vect{y}_{1:I}|m, \vect{d}_{1:I})\pi_{0}(m)}{\sum_{k=1}^{K}\tilde{f}(\vect{y}_{1:I}|k, \vect{d}_{1:I})\pi_{0}(k)},
	\end{align*}
	
	where again the $k$ inside the summation represents $M=k$. We can plug this approximation into the model discrimination utility, given in \eqref{eq:A_utilmd}, to obtain an approximated utility.
	
	\newpage
	
	We can approximate the posterior distribution of the model parameters given the model $m$ and the data $\vect{y}_{1:I}$ via a normal distribution, such that
	
	\begin{align} \vect{\theta}_{m}|m, \vect{y}_{1:I},  \vect{d}_{1:I}  \sim \mathrm{N}\left(\hat{\vect{\theta}}_{m}(\vect{y}_{1:I}), \hat{\vect{\Sigma}}_{m}(\vect{y}_{1:I}) \right). 
	\label{eq:A_N}
	\end{align}
	
	Since we are approximating the posterior of $\vect{\theta}_{m}$ with the multivariate normal distribution displayed in \eqref{eq:A_N}, we can additionally consider approximating the prior distribution with a multivariate normal distribution.  The benefit of approximating both prior and posterior in this way is that there is an explicit expression for the KLD between two multivariate normal distributions.  Thus we can approximate the parameter estimation utility as:
	
	\begin{align*}
	&U(\vect{d}_{1:I}, \vect{y}_{1:I},m) \approx\\
	&\frac{1}{2}\left( \mathrm{tr}(\vect{\Sigma}_{1}^{-1}\hat{\vect{\Sigma}}_{m}(\vect{y}_{1:I}) ) + (\vect{\mu}_{1} - \hat{\vect{\theta}}_{m}(\vect{y}_{1:I}))^{T}\vect{\Sigma}_{1}^{-1}(\vect{\mu}_{1} - \hat{\vect{\theta}}_{m}(\vect{y}_{1:I})) - p +\log\left(\frac{\det \vect{\Sigma}_{1}}{\det \hat{\vect{\Sigma}}_{m}(\vect{y}_{1:I}) } \right) \right),
	\end{align*}
	
	where $\vect{\mu}_{1}$ and $\vect{\Sigma}_{1}$ are the mean and covariance matrix of the prior distribution of $\vect{\theta}_{m}$, respectively.  This significantly simplifies the calculation of the parameter estimation utility. As usual, the expected utility is estimated by Monte Carlo integration over joint samples from $m$ and $\vect{y}_{1:I}$. By using the utility approximations described above, this can be carried out in a computationally efficient way. The only remaining objective is to maximise the expected utility over the design space.
	
	\subsubsection*{ACE Algorithm}
	
	To maximise the expected utility in a high dimensional design space, we use the approximate coordinate exchange (ACE) algorithm \citep{overstall2017bayesian}. Very briefly, this methodology is a cyclic descent algorithm which maximises the expected utility for each design point in turn in a series of optimisation steps. A Gaussian process emulator is used to approximate the expected utility at each optimisation step in order to reduce the number of expected utility calculations required. For a more detailed explanation of the algorithm and the theory behind it, see \citet{overstall2017bayesian}.	
	
	\newpage
	
	\section*{Appendix B}
		
	In this section, we present the results of an additional simulation study. The purpose of this simulation study is to explore the benefit of using the different utility functions for our experimental goals within sequential design.

	\subsection*{Simulation Study Details}
	
		For this simulation study, we have a similar set-up to the simulation study described in Section \ref{sec:simstudy} of the main text. However, instead of comparing the performance of sequential and static optimal designs for the different utility functions, we are comparing the performance of different utility functions on our experimental goals. We explore four different methods of selecting the next design point: generate the design points randomly using a uniform distribution over the discrete design space or find the optimal design with respect to one of the three optimal design utilities defined in Section \ref{sec:utilities}.  

For each of the 4 models in Table 1 of the main paper, we generate 10 ``true parameter settings'' from the posterior distribution based on the dataset from \citet{Pap_etal16} (SMC was used to sample from the posterior). Therefore, we have a total of 40 different true parameter/model configurations for this study. For each of these true data generating configurations, we repeat each optimal design algorithm 30 times.  The total number of observations collected in each run of these algorithms is $I = 25$.

For this simulation study, we consider the total exposure time of prey and predator to be $\tau = 24$ hours. The design points $N_{0,i}$ are members of the set $D = \{1,\ldots,300\}$ and the responses can assume the values $n_{i} \in \{0,\ldots,N_{0,i}\}$ for $i = 1,\ldots,I$. The prior distribution of the parameters is given by $\log(a) \sim \mbox{N}(-1.4, 1.35^{2})$, $\log(T_{h}) \sim \mbox{N}(-1.4, 1.35^{2})$ and $\log(\lambda) \sim \mbox{N}(-1.4, 1.35^{2})$. The parameter $\lambda$ is only required in the beta-binomial models. In addition, the prior model probabilities are equal across the $K=4$ models.

We use the Bayesian D-posterior precision to quantify the precision of the true model parameters. By comparing this across different design selection methods, we can identify which method is the most efficient for parameter estimation. In a similar way, comparing the model probabilities across different design selection methods enables us to compare and assess the model discrimination power of each method. The subsequent section displays the results of the simulation study.

\subsection*{Results}

\begin{figure}[!htp]
	\centering
	\subfigure[True Model 1]{\includegraphics[height=0.193\textheight,keepaspectratio]{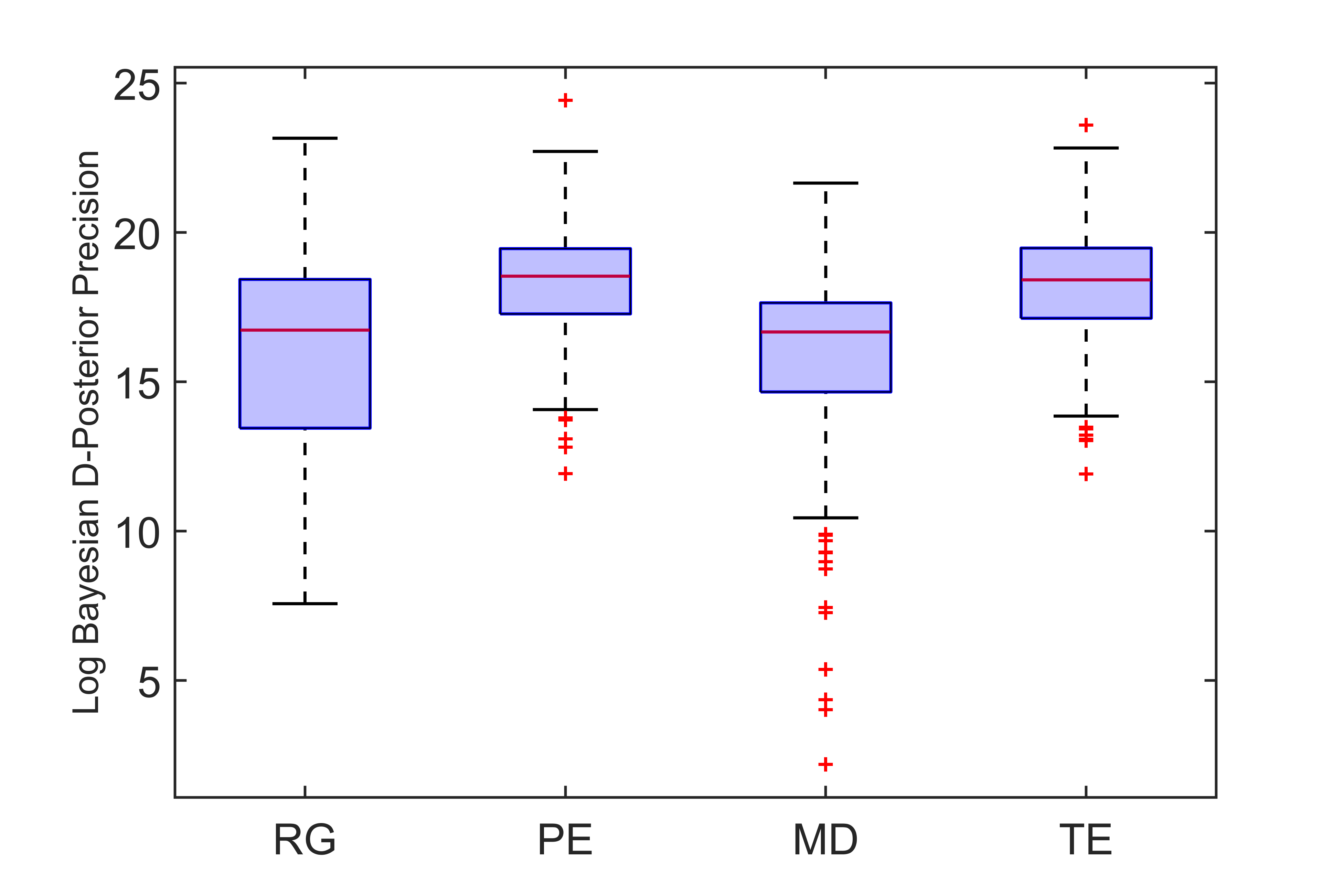}\label{figsub:pea}}
	\subfigure[True Model 2]{\includegraphics[height=0.193\textheight,keepaspectratio]{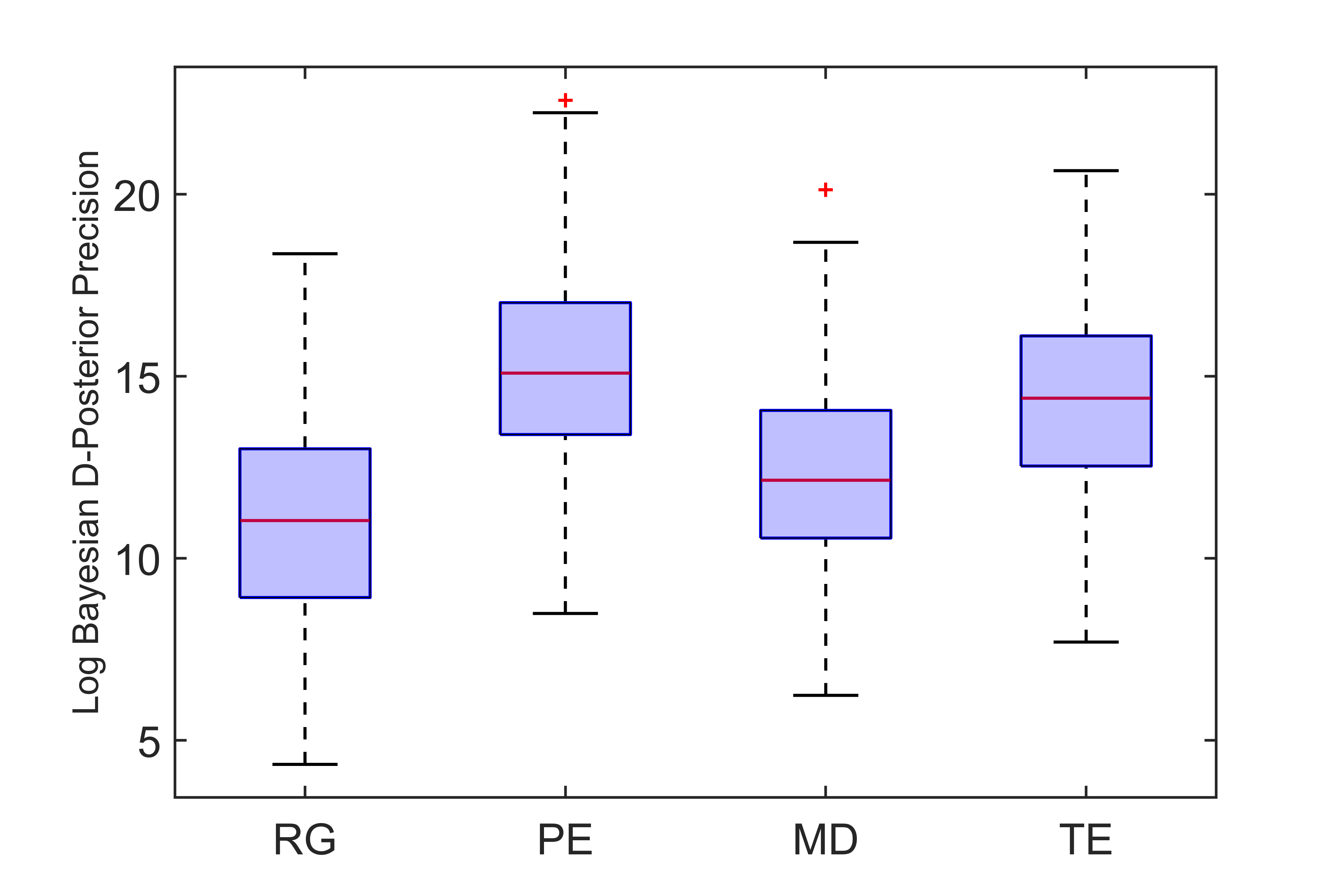}\label{figsub:peb}}
	\subfigure[True Model 3]{\includegraphics[height=0.193\textheight,keepaspectratio]{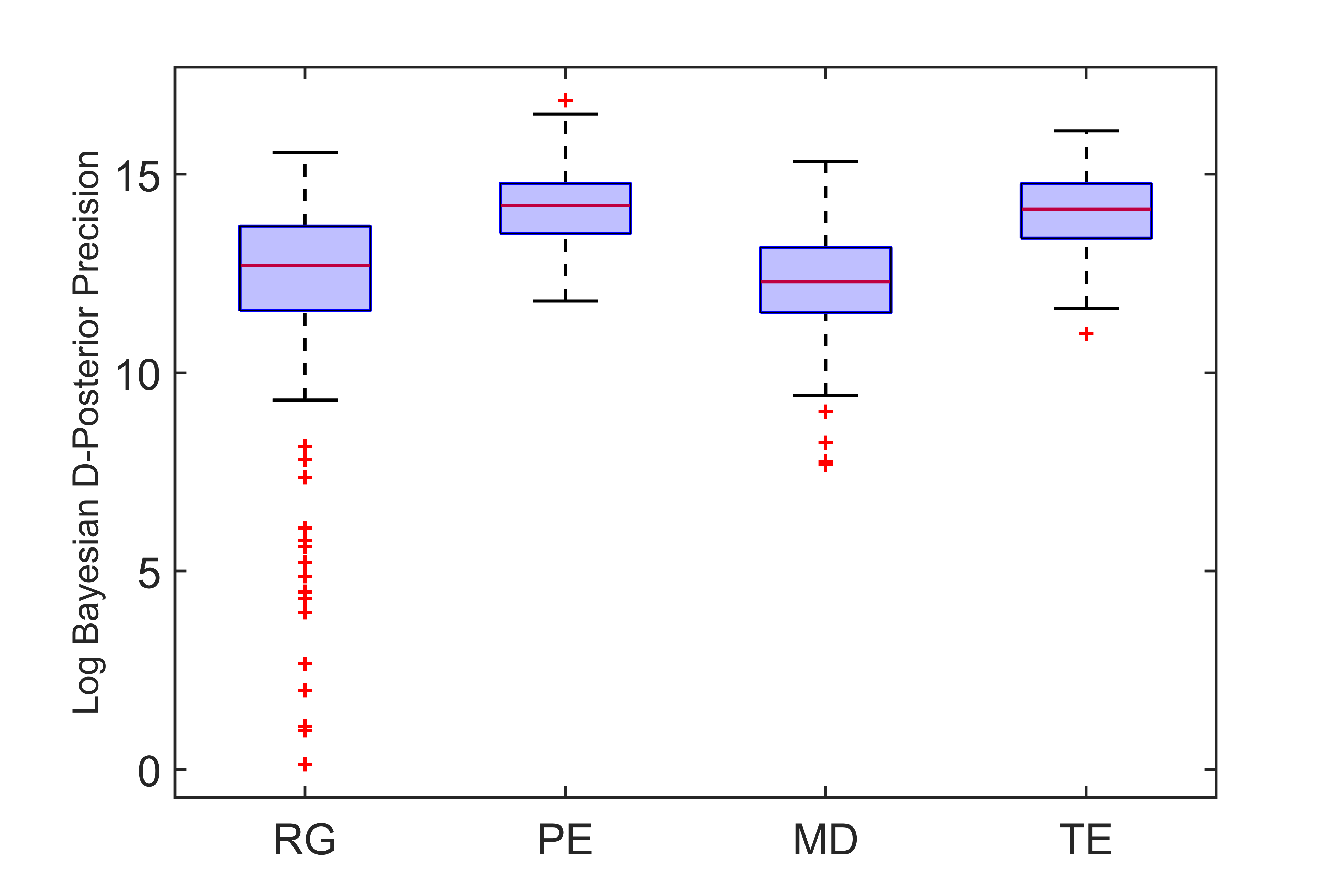}\label{figsub:pec}}
	\subfigure[True Model 4]{
		\includegraphics[height=0.193\textheight,keepaspectratio]{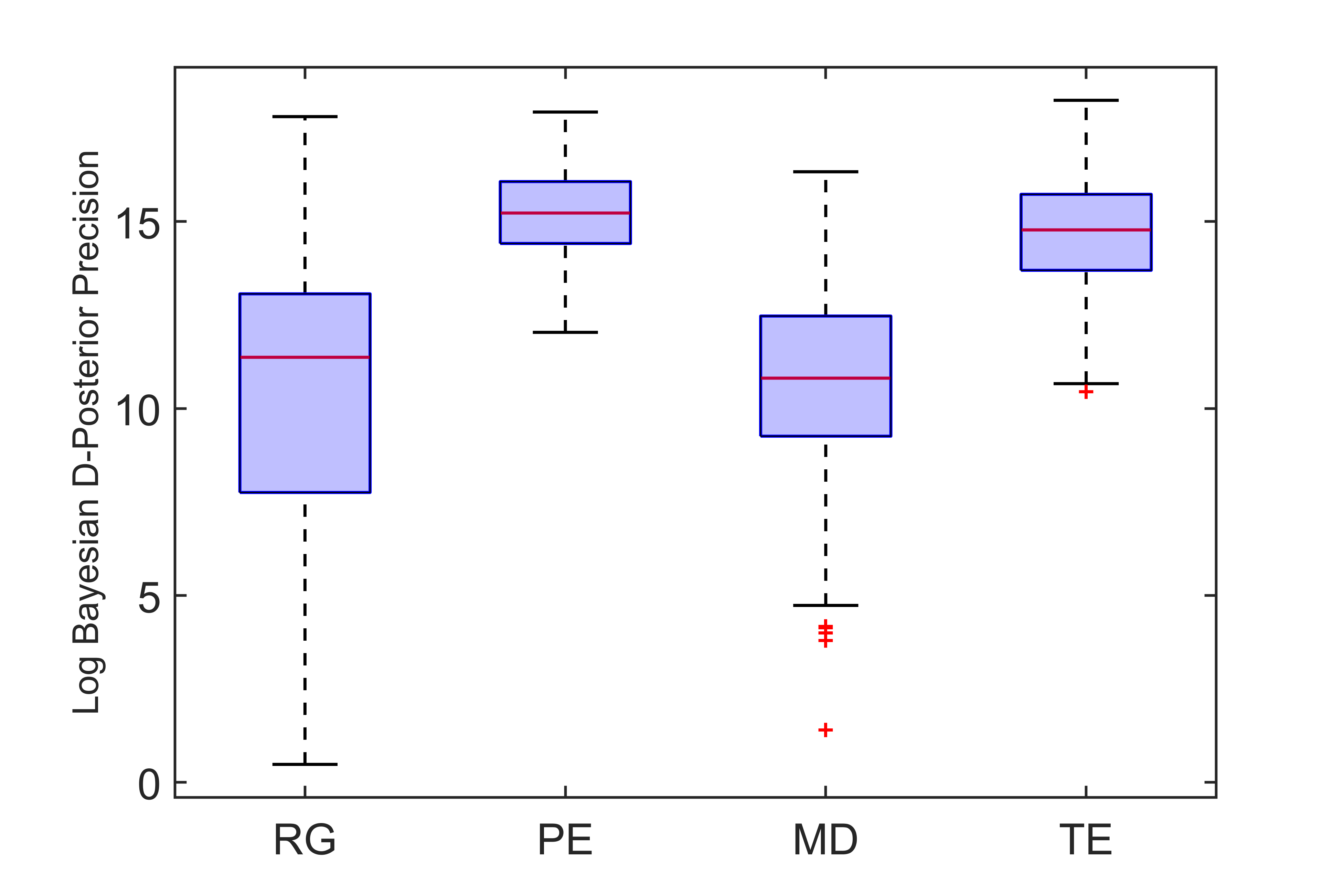}\label{figsub:ped}}
	\caption{Distribution of log Bayesian D-posterior precision of the true model across design selection methods and different true models. RG, PE, MD, and TE represent randomly generated, parameter estimation, model discrimination and dual-purpose (total entropy) design selection methods, respectively.}
	\label{fig:pe}
\end{figure}

We begin our analysis of the simulation study results by examining the precision of our parameters in the true model after the SMC design process. Figure \ref{fig:pe} displays the distribution of the log Bayesian D-posterior precision at experiment number $I=25$ across the different design selection methods for each true model.  It is apparent from Figure \ref{fig:pe} that there is a noticeable pattern in the results across the different true models. The parameter estimation design selection method outperforms the other methods, as expected. However, the total entropy (which incorporates model discrimination and parameter estimation) only performs marginally worse overall compared with the parameter estimation design. Furthermore, we find the model discrimination design does not perform well for parameter estimation. 

\newpage

\begin{figure}[!htp]
	\centering
	\subfigure[True Model 1]{\includegraphics[height=0.19\textheight,keepaspectratio]{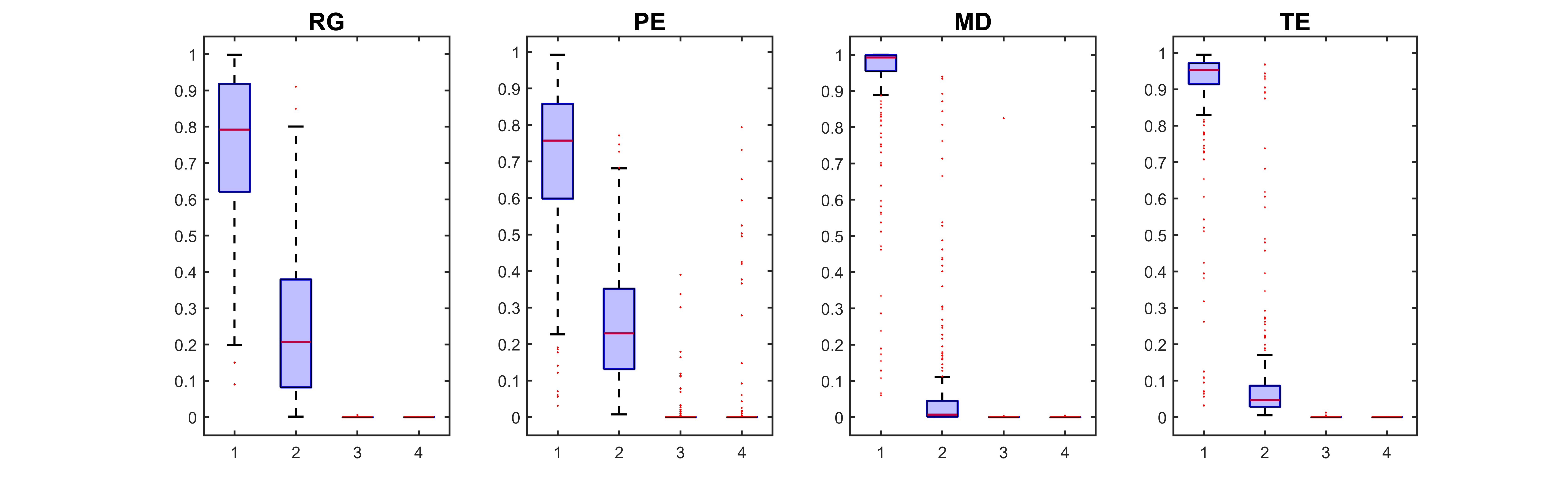}\label{figsub:mda}}
	\subfigure[True Model 2]{\includegraphics[height=0.19\textheight,keepaspectratio]{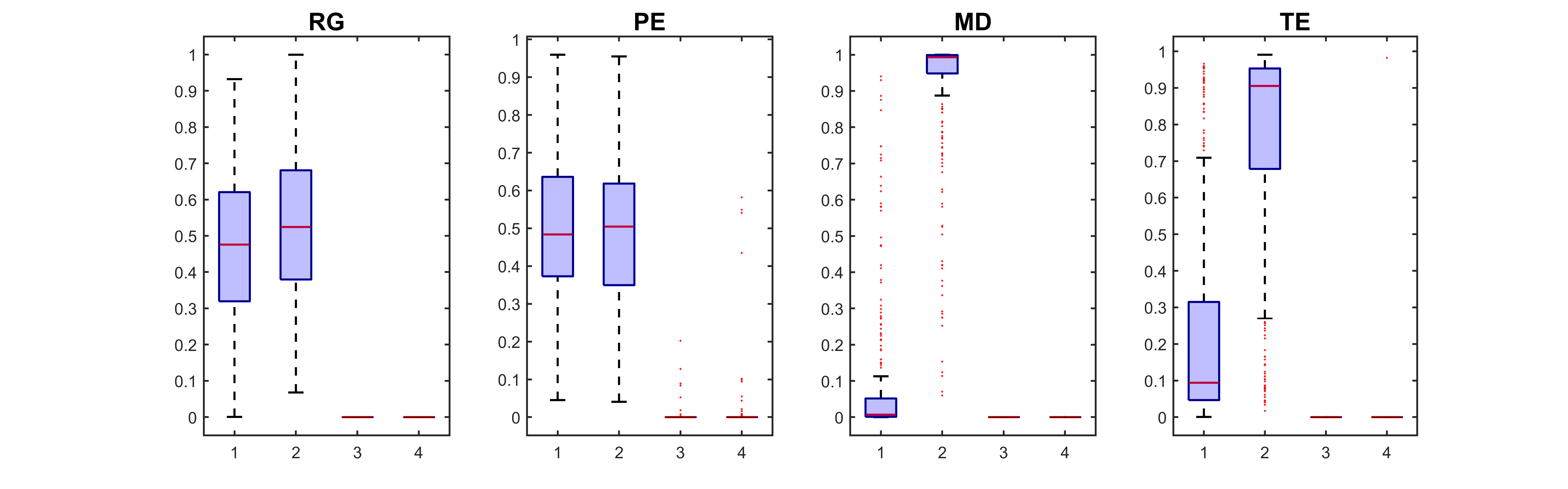}\label{figsub:mdb}}
	\subfigure[True Model 3]{\includegraphics[height=0.19\textheight,keepaspectratio]{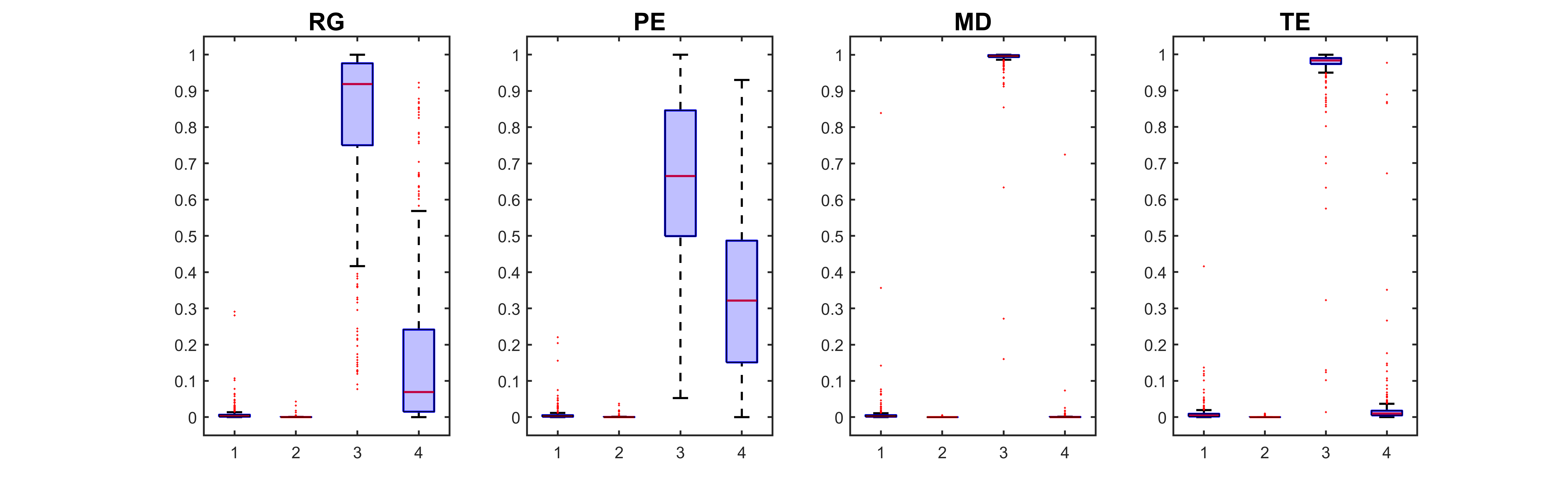}\label{figsub:mdc}}
	\subfigure[True Model 4]{
		\includegraphics[height=0.19\textheight,keepaspectratio]{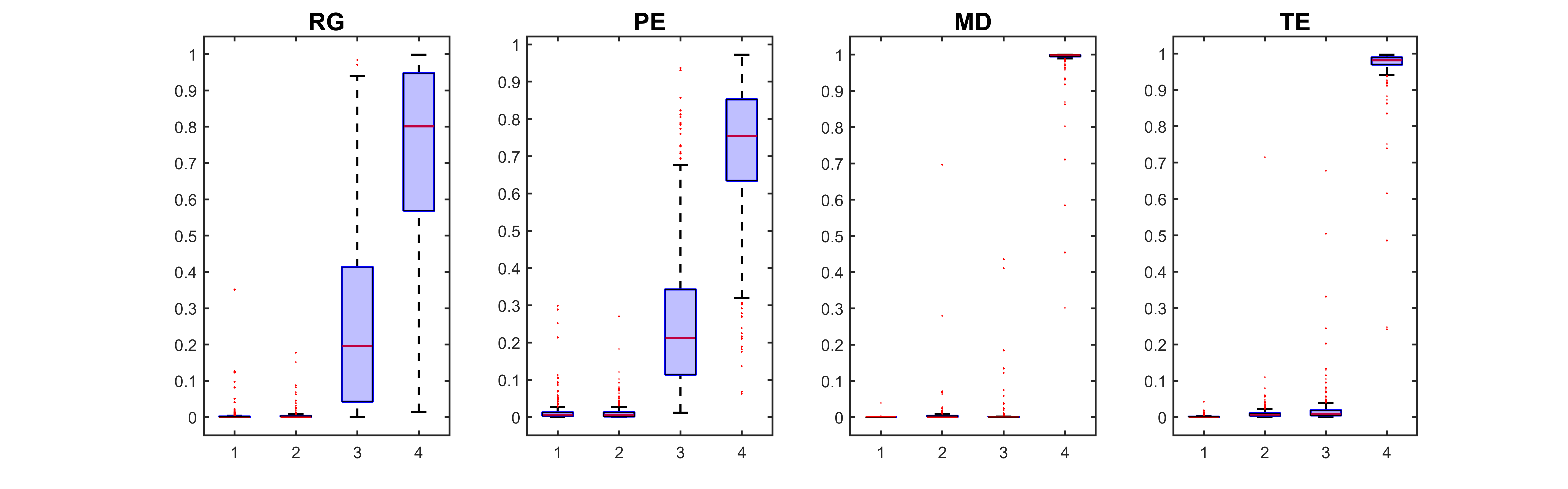}\label{figsub:mdd}}
	\caption{Distribution of posterior model probabilities across design selection methods for different true models. RG, PE, MD, and TE represent randomly generated, parameter estimation, model discrimination and dual-purpose (total entropy) design selection methods, respectively. The x-axis on each plot represents the model number and corresponds to the model numbers in Table 1. The y-axis shows the posterior model probability.}
	\label{fig:md}
\end{figure}

The distributions of posterior model probabilities at experiment number $I=25$ for the different design selection methods and each true model are shown in Figure \ref{fig:md}. As we would expect, the model discrimination design selection method outperforms the other methods for all true models. The dual-purpose design is only marginally less effective. It is evident from Figures \ref{figsub:mda} and \ref{figsub:mdb} that for the randomly generated and parameter estimation design selection methods, the algorithm struggles to discriminate between the beta-binomial models when the true model is beta-binomial. Similarly, Figures \ref{figsub:mdc} and \ref{figsub:mdd} show that for the same two design selection methods, the SMC design algorithm does not effectively discriminate between the binomial models when the true model is binomial. The parameter estimation design selection method does not perform well for the goal of model discrimination.

In both experimental goals the total entropy utility is only marginally less effective than the single-purpose utilities. This suggests that this utility is beneficial for efficient parameter estimation and model discrimination simultaneously. 

\begin{figure}[!htp]
	\centering
	\subfigure[True Model 1]{\includegraphics[height=0.193\textheight,keepaspectratio]{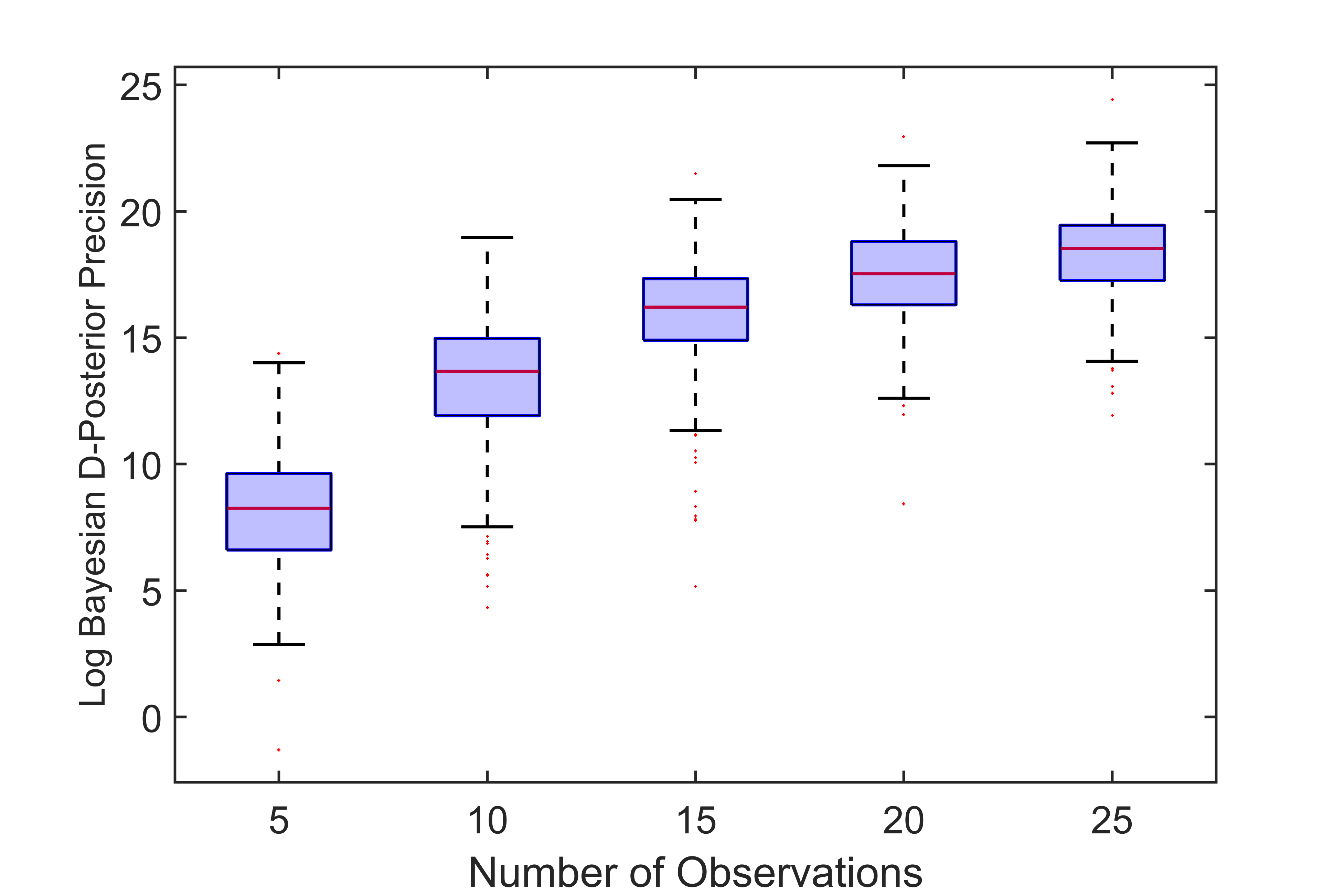}\label{figsub:peoa}}
	\subfigure[True Model 2]{\includegraphics[height=0.193\textheight,keepaspectratio]{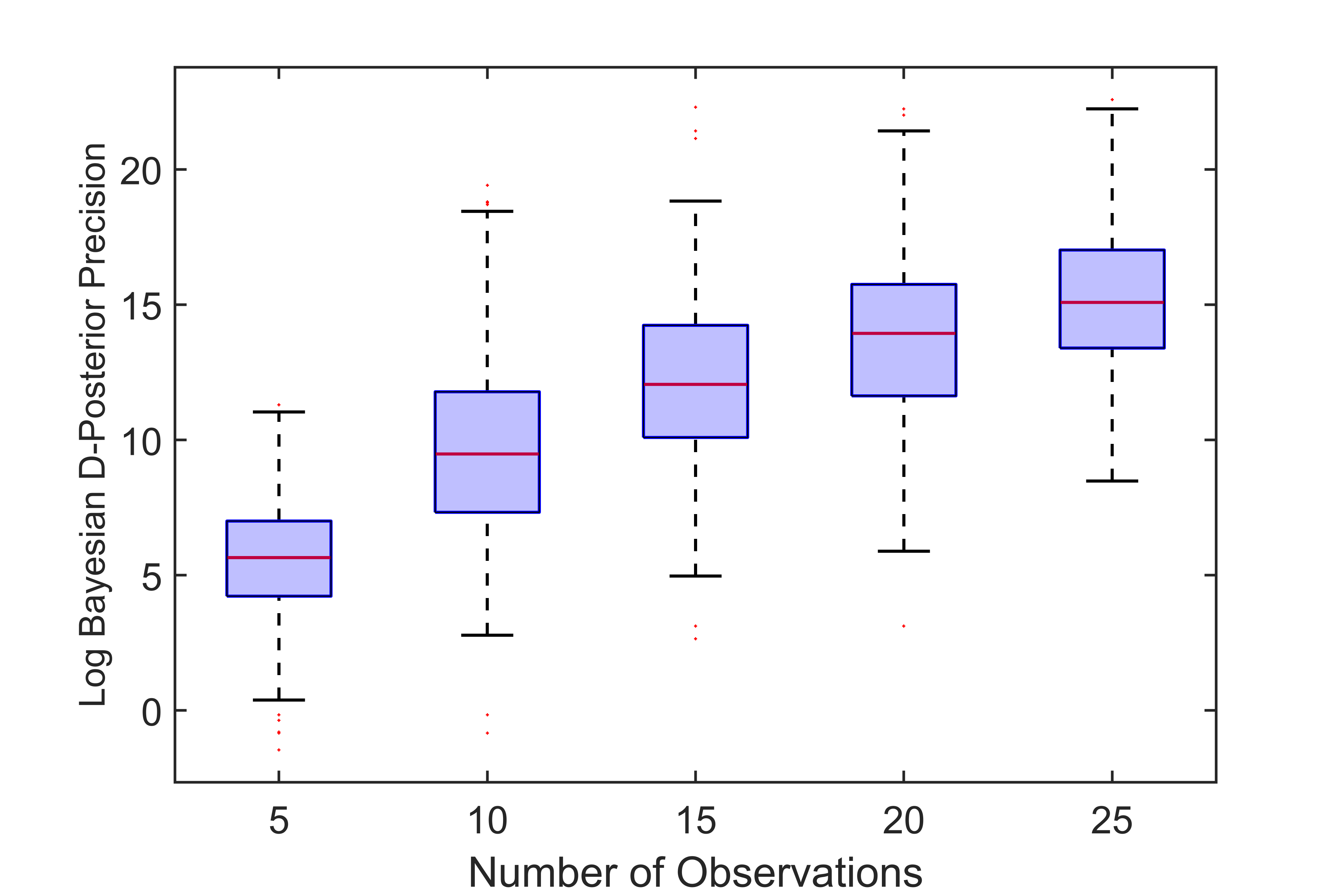}\label{figsub:peob}}
	\subfigure[True Model 3]{\includegraphics[height=0.193\textheight,keepaspectratio]{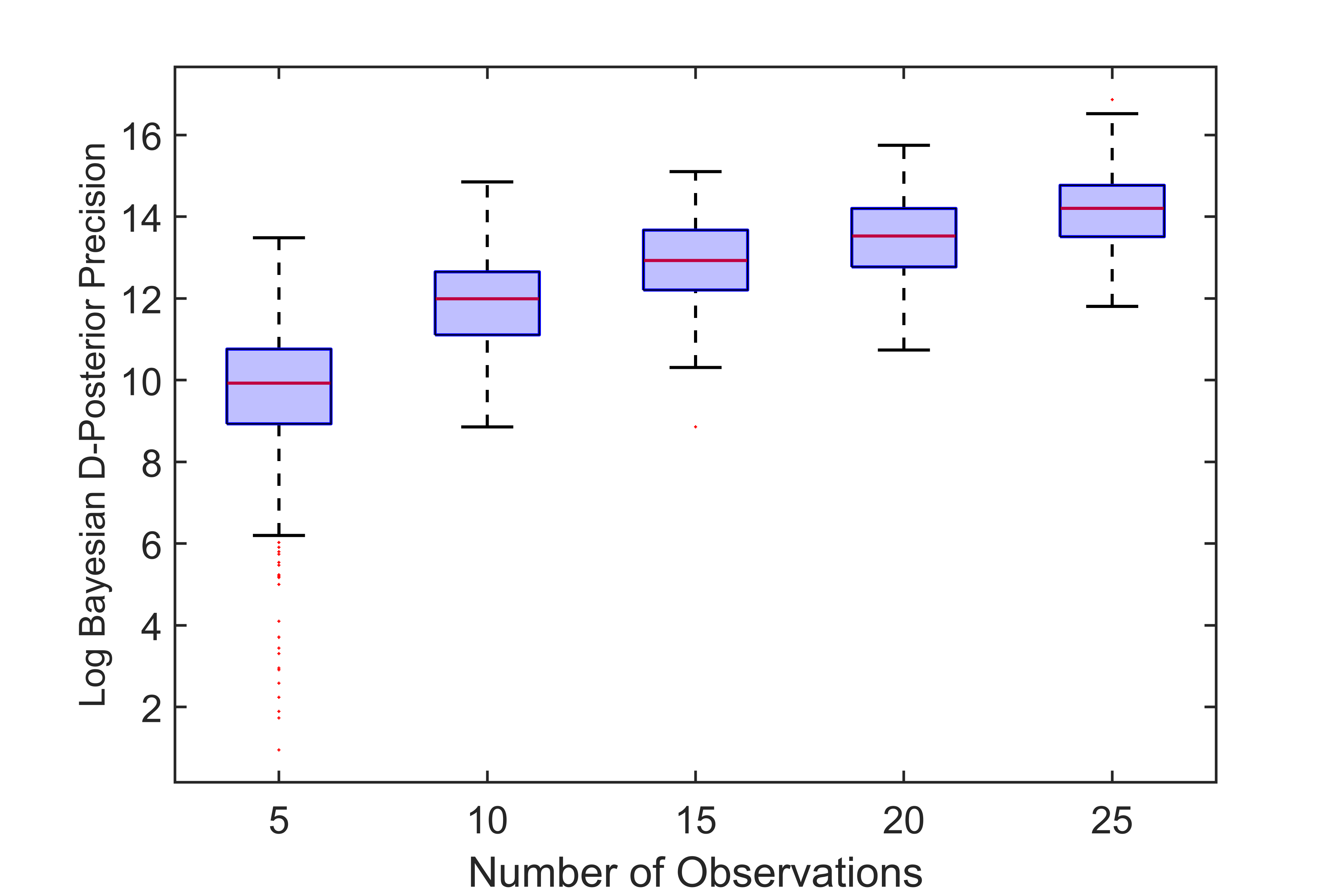}\label{figsub:peoc}}
	\subfigure[True Model 4]{
		\includegraphics[height=0.193\textheight,keepaspectratio]{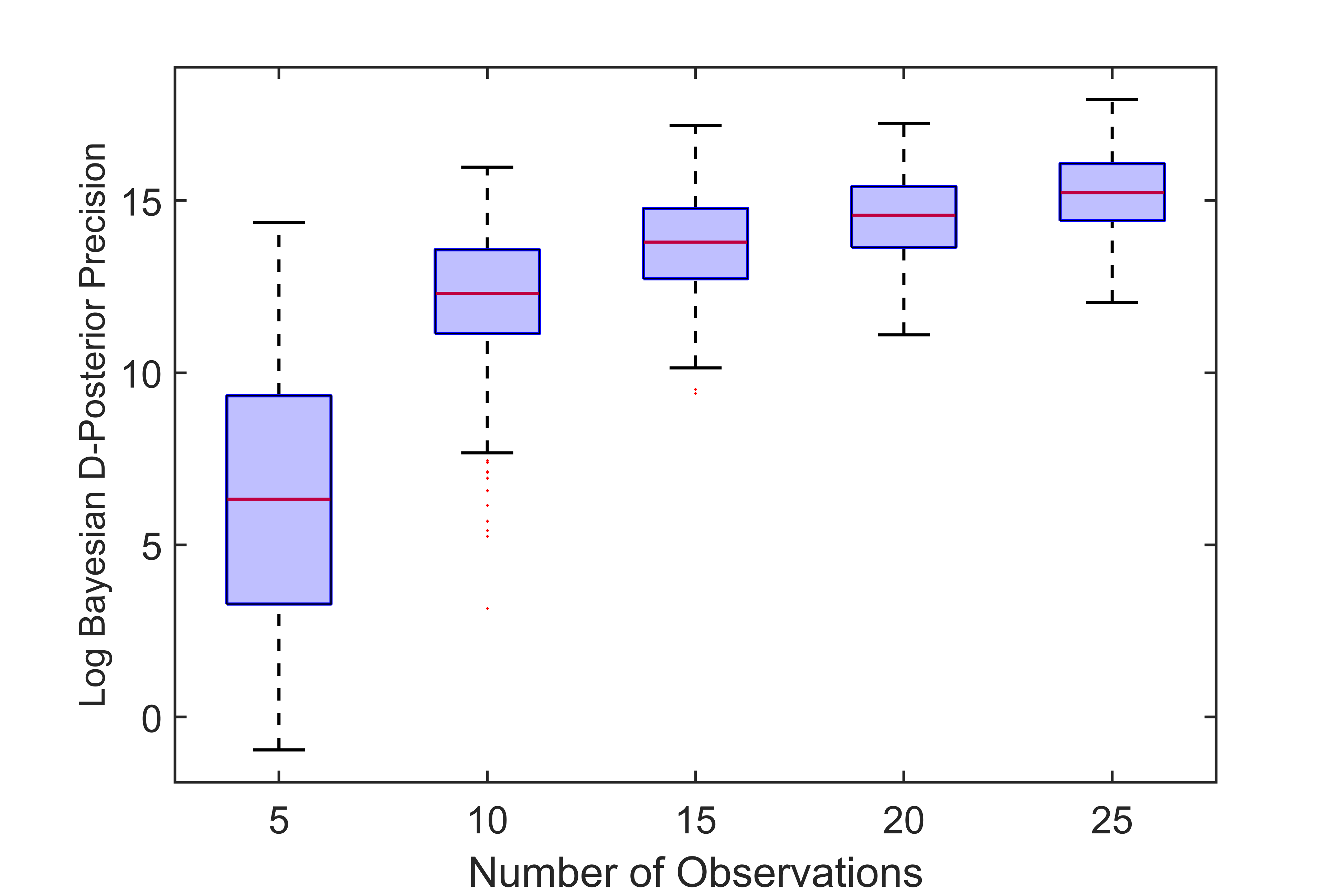}\label{figsub:peod}}
	\caption{Distribution of log Bayesian D-posterior precision across multiple iterations of the SMC design process for different true models. Results are generated from running SMC using the parameter estimation design selection method. The x-axis on each plot represents the number of experiments that have currently been run (i.e. the number of iterations of the SMC algorithm or number of observations collected).}
	\label{fig:peo}
\end{figure}

Figure \ref{fig:peo} displays the distribution of the log Bayesian D-posterior precision for each true model at several iterations in the SMC design process. This enables us to identify the experimental number at which the gain in parameter precision becomes negligible. The results in Figure \ref{fig:peo} are generated from running SMC using the parameter estimation design selection method. Figure \ref{fig:peo} suggests that the increase in precision for all true models seems to decay exponentially.

\begin{figure}[!htp]
	\centering
	\subfigure[True Model 1]{\includegraphics[height=0.18\textheight,keepaspectratio]{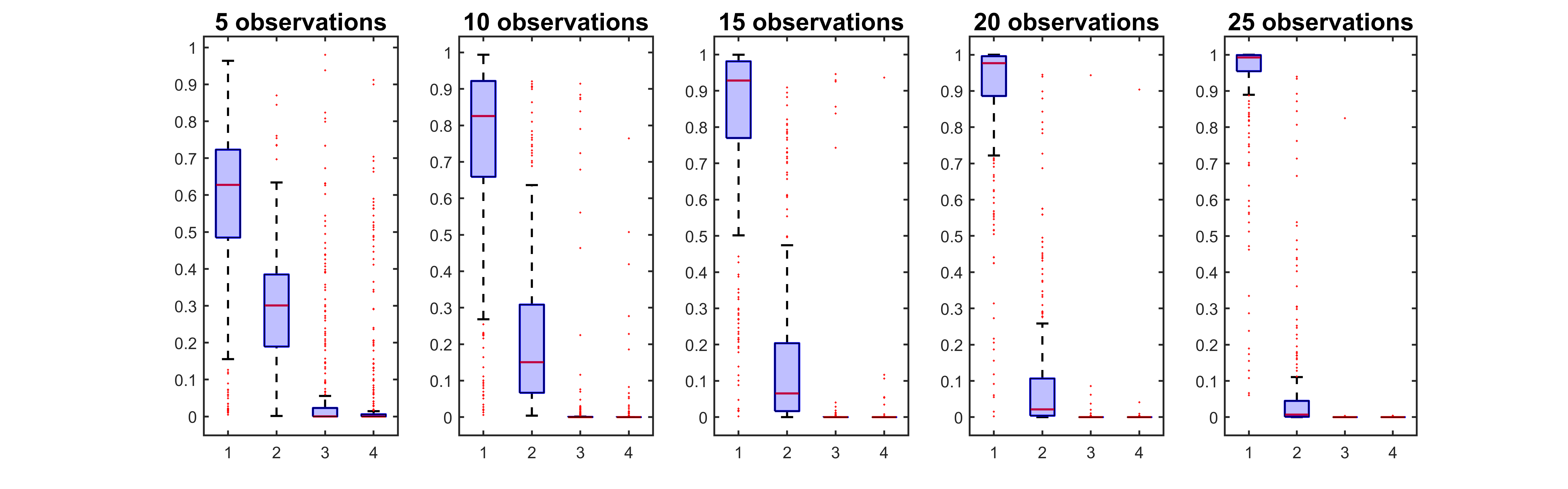}\label{figsub:mdoa}}
	\subfigure[True Model 2]{\includegraphics[height=0.18\textheight,keepaspectratio]{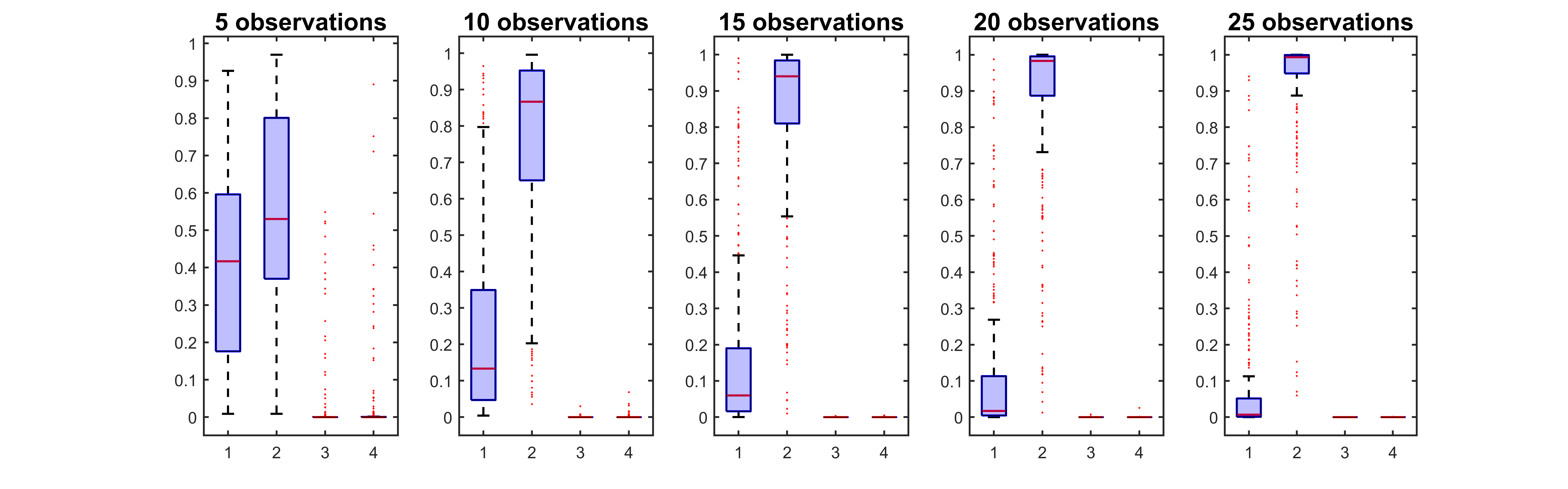}\label{figsub:mdob}}
	\subfigure[True Model 3]{\includegraphics[height=0.18\textheight,keepaspectratio]{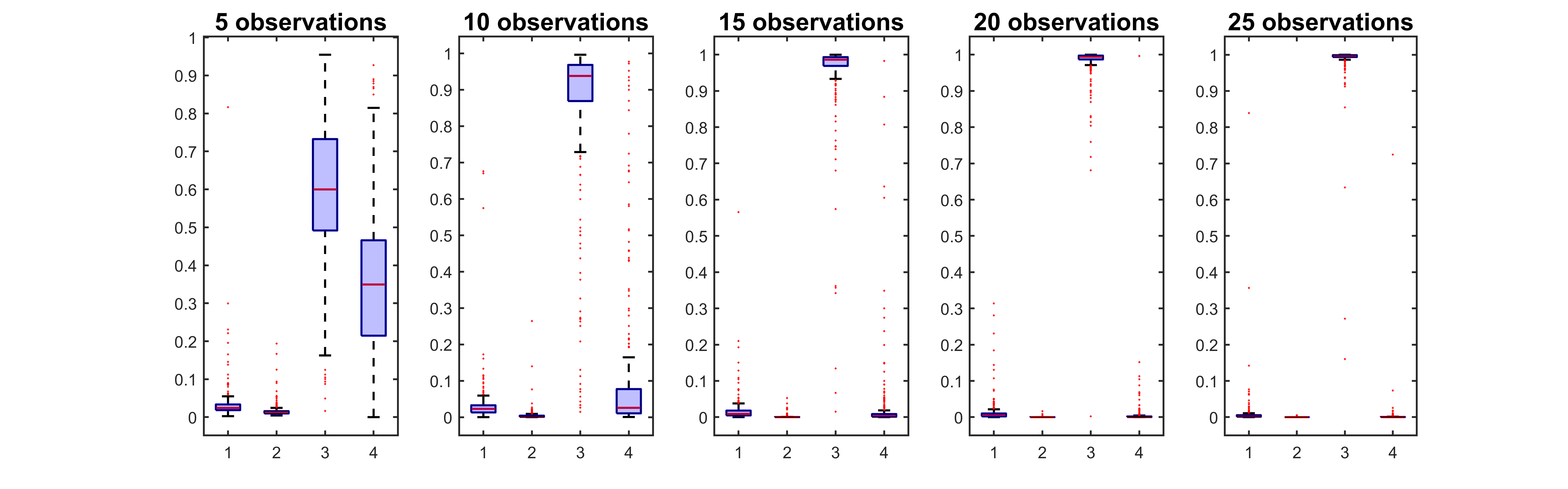}\label{figsub:mdoc}}
	\subfigure[True Model 4]{
		\includegraphics[height=0.18\textheight,keepaspectratio]{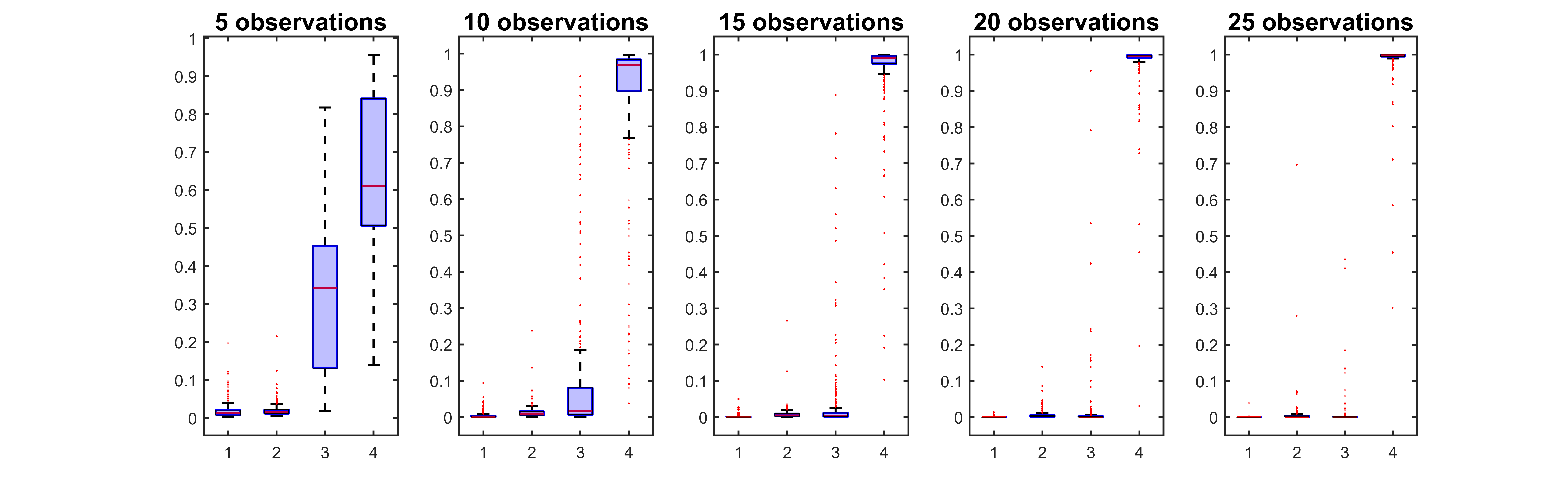}\label{figsub:mdod}}
	\caption{Distribution of posterior model probabilities across multiple iterations of the SMC design process for different true models.  Results are generated from running SMC using the model discrimination design selection method. The x-axis on each plot represents the model number and corresponds to the model numbers in Table 1. The y-axis represents the posterior model probability. The titles on each plot represent the number of experiments that have currently been run (i.e. the number of iterations of the SMC algorithm or number of observations collected). }
	\label{fig:mdo}
\end{figure}

Figure \ref{fig:mdo} shows the posterior model probabilities after 5, 10, 15, 20 and 25  experiments have been run. The results in this figure are generated using the model discrimination design selection method.  Figure \ref{fig:mdo} enables us to identify how much gain in model discriminative ability can be achieved with increasing the sample size for each true model.  It is evident that after 25 experiments, Models 3 and 4 can already be identified with high probability.

\begin{figure}[!htp]
	\centering
	\subfigure[Parameter Estimation Design Selection]{\includegraphics[height=0.194\textheight,keepaspectratio]{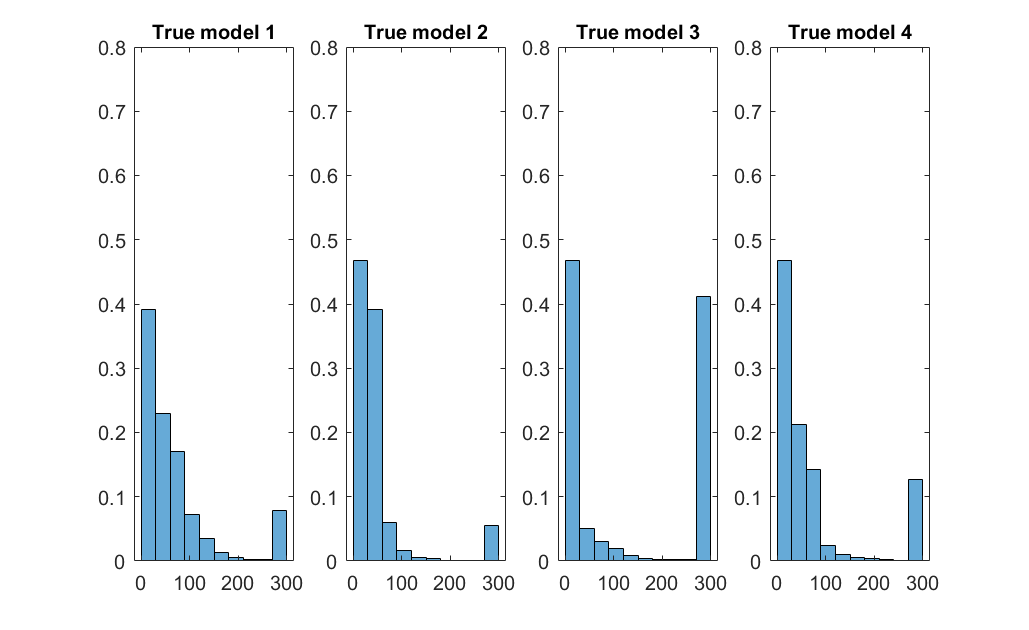}\label{figsub:dpa}}
	\subfigure[Model Discrimination Design Selection]{\includegraphics[height=0.194\textheight,keepaspectratio]{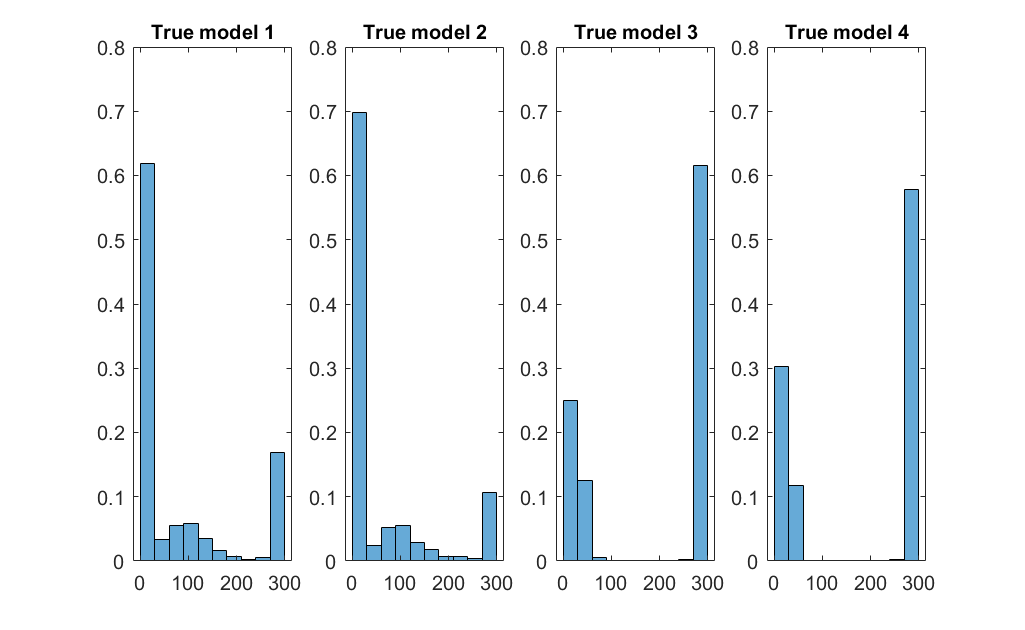}\label{figsub:dpb}}
	\subfigure[Dual-Purpose Design Selection]{
		\includegraphics[height=0.194\textheight,keepaspectratio]{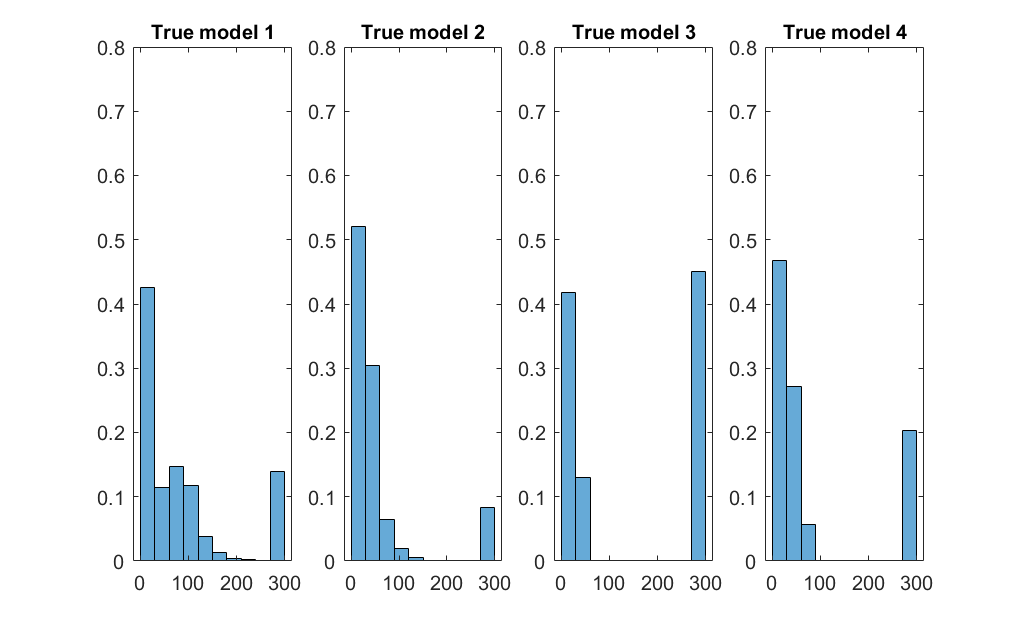}\label{figsub:dpc}}
	\caption{Distribution of design points after $I=25$ experiments for the different true models and the different design selection methods. The x-axis of each plot represents the design points and the y-axis represents the relative frequency. The titles of the plots indicate which model is the true model.}
	\label{fig:dp}
\end{figure}

The distributions of design points for the different utility functions and different true models are displayed in Figure \ref{fig:dp}. Analysing these distributions separately provides insight into where we need to place design points to gain the most information about parameters and the true model.  The designs for our three proposed design selection methods are predominantly bimodal with modes at boundaries of the design space.  Therefore, it appears selecting design points at the boundaries is optimal for parameter estimation and model discrimination. The designs for parameter estimation are less concentrated in the lower third of the design space than the model discrimination designs. However, the parameter estimation designs are still heavily right-skewed in the lower third. The design point distribution for the dual-purpose design selection method seems to contain features of both the parameter estimation and model discrimination design distributions. This is expected as the dual-purpose utility is formed from the other two utilities. The model discrimination designs differ quite strongly between the binomial and beta-binomial models as shown in Figure \ref{fig:dp}. The beta-binomial designs have comparatively little mass at the upper boundary (almost all mass at the lower boundary), the binomial designs have much more mass at the upper boundary.

\newpage

\end{document}